\documentclass[sigconf]{acmart}

\AtBeginDocument{%
  }

\copyrightyear{2025}
\acmYear{2025}
\setcopyright{acmlicensed}\acmConference[MM '25]{Proceedings of the 33rd
ACM International Conference on Multimedia}{October 27--31, 2025}{Dublin,
Ireland}
\acmBooktitle{Proceedings of the 33rd ACM International Conference on
Multimedia (MM '25), October 27--31, 2025, Dublin, Ireland}
\acmDOI{10.1145/3746027.3755502}
\acmISBN{979-8-4007-2035-2/2025/10}

\usepackage{pifont}
\usepackage{algorithm}
\usepackage{algorithmic}

\usepackage{amssymb}
\usepackage{amsmath}
\usepackage{amsfonts}
\usepackage{footmisc}
\usepackage{multirow}
\usepackage{booktabs}
\usepackage[utf8]{inputenc}
\usepackage{url}
\usepackage{bbding}
\usepackage{wasysym}
\usepackage{utfsym}
\usepackage{xcolor,soul}
\usepackage{fontawesome}
\usepackage{hyperref}
\usepackage{array}
\usepackage{svg}
\usepackage{tikz}
\usepackage{circledsteps}
\usepackage{colortbl}
\usepackage[most]{tcolorbox}
\usepackage{lipsum}
\usepackage{enumitem}

\begin{document}

\title{UniTalker: Conversational Speech-Visual Synthesis}


\author{Yifan Hu}
\email{22309013@mail.imu.edu.cn}
\orcid{0009-0008-2276-1456}
\affiliation{
  \institution{Inner Mongolia University}
  \city{Hohhot}
  \country{China}
}

\author{Rui Liu}
\authornote{Corresponding author.}
\email{imucslr@imu.edu.cn}
\orcid{0000-0003-4524-7413}
\affiliation{
  \institution{Inner Mongolia University}
  \city{Hohhot}
  \country{China}
}

\author{Yi Ren}
\email{ren.yi@bytedance.com}
\orcid{0000-0002-9160-3848}
\affiliation{
  \institution{ByteDance}
  \city{Singapore}
  \country{Singapore}
}

\author{Xiang Yin}
\email{yinxiang.stephen@bytedance.com}
\orcid{0000-0003-0472-2783}
\affiliation{
  \institution{ByteDance}
  \city{Shanghai}
  \country{China}
}

\author{Haizhou Li}
\email{haizhouli@cuhk.edu.cn}
\orcid{0000-0001-9158-9401}
\affiliation{
  \department{SRIBD, School of Data Science}
  \institution{The Chinese University of Hong Kong}
 \city{Shenzhen}
 \country{China}
}

\renewcommand{\shortauthors}{Yifan Hu, Rui Liu, Yi Ren, Xiang Yin, and Haizhou Li.}

\begin{abstract}
Conversational Speech Synthesis (CSS) is a key task in the user-agent interaction area, aiming to generate more expressive and empathetic speech for users. 
However, it is well-known that “listening” and “eye contact” play crucial roles in conveying emotions during real-world interpersonal communication.
Existing CSS research is limited to perceiving only text and speech within the dialogue context, which restricts its effectiveness. Moreover, speech-only responses further constrain the interactive experience. 
To address these limitations, we introduce a Conversational Speech-Visual Synthesis (CSVS) task as an extension of traditional CSS.
By leveraging multimodal dialogue context, it provides users with coherent audiovisual responses. To this end, we develop a CSVS system named \textbf{UniTalker}, which is a unified model that seamlessly integrates multimodal perception and multimodal rendering capabilities. 
Specifically, it leverages a large-scale language model to comprehensively understand multimodal cues in the dialogue context, including speaker, text, speech, and the talking-face animations. 
After that, it employs multi-task sequence prediction to first infer the target utterance's emotion and then generate empathetic speech and natural talking-face animations. 
To ensure that the generated speech-visual content remains consistent in terms of emotion, content, and duration, we introduce three key optimizations:
1) Designing a specialized neural landmark codec to tokenize and reconstruct facial expression sequences.
2) Proposing a bimodal speech-visual hard alignment decoding strategy.
3) Applying emotion-guided rendering during the generation stage.
Comprehensive objective and subjective experiments demonstrate that our model synthesizes more empathetic speech and provides users with more natural and emotionally consistent talking-face animations. The source code and generated samples are available at: \url{https://github.com/AI-S2-Lab/UniTalker}.
\end{abstract}




\maketitle

\begin{figure}[t]
\centering
\centerline{
\includegraphics[width=0.9\linewidth]{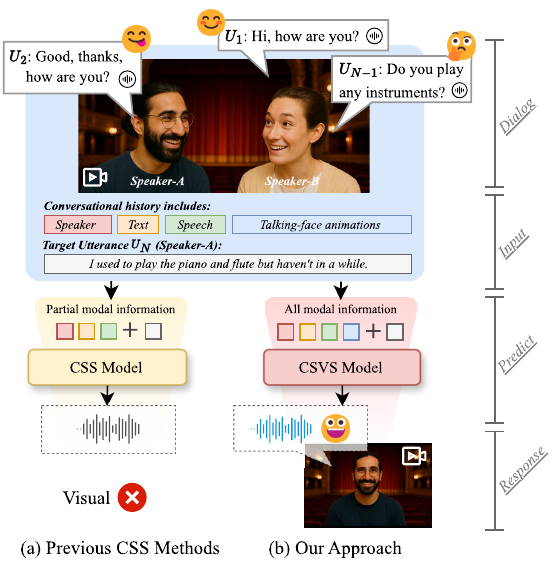}
}
\caption{(a) Previous CSS methods synthesize the target speech from partial dialogue context.
(b) Our approach leverages full-modal dialogue context to predict emotion, speech, and talking-face animations, enabling both auditory and visual feedback.}
\label{fig:case}
\vspace{-2mm}
\end{figure}
\section{Introduction}
Traditional conversational speech synthesis (CSS) systems aim to generate expressive speech that aligns with the current dialogue context, as shown in Fig. \ref{fig:case} (a). For example, ~\cite{xue2023m2ctts,li2022inferring, hu2024fctalker} employ various encoders to extract multi-scale textual and speech features from the context to establish local and global dependencies in context modeling. ~\cite{li2022inferring} and ~\cite{liu2024emotion} utilize graph neural networks to establish relationships between different utterances within the context. Meanwhile, ~\cite{liu2024gpttalker} leverages the powerful contextual reasoning capabilities of large language models (LLMs) to directly predict the speech token sequences, achieving improvements in naturalness and expressiveness. 
However, these previous CSS systems respond to users with purely speech, resulting in monotonous content that limits user interaction experiences in scenarios such as virtual reality interactions \cite{jerald2017vr} and virtual digital humans \cite{bajaj2024design}. Moreover, in real-world human communication, individuals not only understand each other's emotions through speech but also interpret facial expressions visually, enabling perception and responses based on multimodal conversational cues. To address this, we extend the traditional CSS task to conversational speech-visual synthesis. Specifically, the constructed models need to integrate visual information into the input context and simultaneously generate both speech and talking-face animations as outputs, as shown in Fig. \ref{fig:case} (b). The task goal remains to ensure that the multimodal responses align with the current interaction context. This raises a key question: how to make effective use of visual information and generate appropriate responses in CSVS?

Many studies now highlight the importance of visual modalities in user-agent interaction and integrate them into models to aid understanding and generation. For example, in dialogue response generation tasks, ~\cite{fei2024empathyear} directly uses the synthesized speech to drive a digital portrait in a cascading manner. However, when the synthesized speech signal is weak, it leads to instability in the generation process. ~\cite{zhang2025empatheia} employs LLM to directly predict video-driving intermediate representations, incorporating global emotional and stylistic information to guide the generator in producing talking-face animations. Although this approach improves the emotional consistency of the responses, it lacks finer facial control.
In speech-driven portrait animation tasks, some recent work explores using facial keypoints (also called landmarks) to assist the driving process, achieving more expressive and natural results \cite{chen2024echomimic,wei2024aniportrait}. Nevertheless, the keypoint information often comes from manual input or is directly predicted from speech, lacking a generation method better suited for the CSVS task to ensure that facial expressions match the dialogue context.

To address these challenges, we develop \textbf{UniTalker}, a powerful CSVS system with multimodal dialogue context awareness and speech-visual generation capabilities. Specifically, to ensure more stable talking-face animations, we adopt a joint speech–facial expression driving strategy, where facial landmark sequences guide the animation alongside speech.  
Inspired by the strong contextual reasoning and multitask prediction capabilities of LLMs \cite{kim2024unified, lin2024paralinguistics, pang2024masked, chen2024large}, we adopt an LLM-based module, named \textbf{EVSLM}, to directly predict the facial landmarks and speech of the target utterance based on the dialogue context. To ensure that the predicted facial expression sequence and speech are aligned in both duration and content (i.e., synchronizing lip movements with spoken content), we design a special neural codec to tokenize and reconstruct facial landmarks. Notably, both the codec and the speech tokenizer operate under frame-level modeling with the same token rate, which naturally facilitates alignment between facial expressions and speech. 
Based on this design, EVSLM alternately predicts the speech and visual tokens of the target utterance during contextual modeling. Furthermore, to ensure emotional consistency, EVSLM first predicts the target utterance's emotion from the dialogue context, then guides the speech renderer to synthesize empathetic speech. The predicted facial landmarks and speech are subsequently used to render emotionally consistent and natural talking-face animations.

Overall, our contributions are as follows: \textbf{1) Unified Multimodal Perception and Rendering:} we develop the UniTalker model, which integrates multimodal context understanding and speech-visual generation capabilities. While responding with empathetic speech, it also generates more natural and emotionally consistent talking-face animations.
\textbf{2) Specialized Low-token-rate Landmark Codec:} We design a special neural codec with a token rate of 25 Hz to more effectively tokenize and reconstruct facial expression sequences in talking-face animations.
\textbf{3) Audiovisual Consistent Response:} We propose a bimodal speech-visual hard alignment strategy. After the EVSLM predicts the target utterance's emotion, the model alternately decodes visual and speech tokens, ensuring token-level alignment in duration and content.
\textbf{4) Comprehensive Evaluation:} Comprehensive objective and subjective experiments demonstrate the superiority of our UniTalker model in the CSVS task, as well as the effectiveness of the proposed landmark codec and bimodal hard alignment decoding strategy.

\begin{figure*}[t]
\centering
\centerline{
\includegraphics[width=0.91\linewidth]{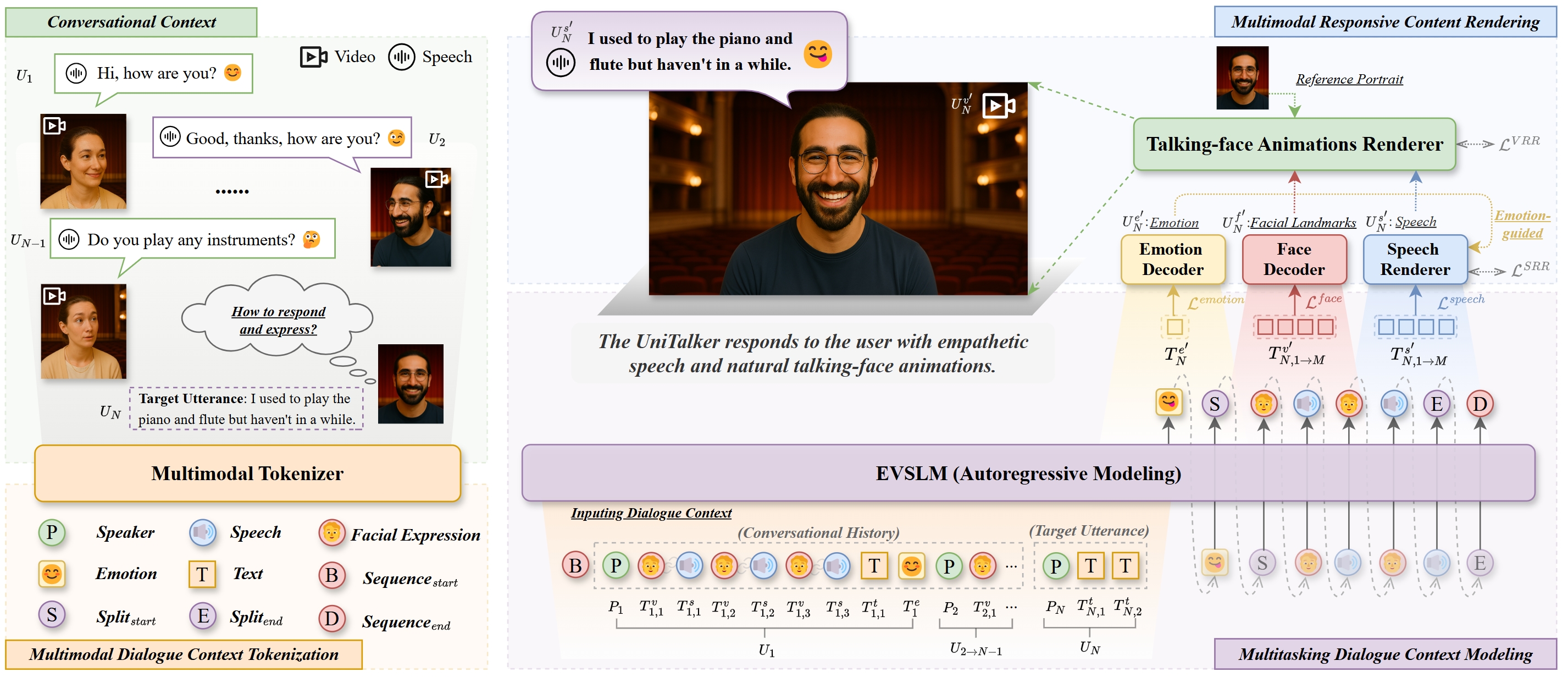}
}
\caption{The overall architecture of UniTalker. The model first repersents multimodal information from conversations through \textit{Multimodal Dialogue Context Tokenization}. Then, during \textit{Multitasking Dialogue Context Modeling}, it interprets these multimodal cues and sequentially predicts token sequences for the target utterance’s emotion, facial expressions, and speech. Finally, through \textit{Multimodal Responsive Content Rendering}, it generates and responds empathetic speech $U_N^{s^{\prime}}$ and natural talking-face animations $U_N^{v^{\prime}}$ to the user.}
\label{fig:model}
\end{figure*}

\section{Related Works}
\subsection{Spoken Dialogue Systems}
LLM-based dialogue systems have made significant strides, initially focusing on understanding multimodal inputs and generating text responses across tasks such as story continuation \cite{yang2022story}, translation \cite{xu2024translation}, code generation \cite{jiang2024code}, visual reasoning \cite{liu2023llava}, and audio analysis \cite{chu2023qwenaudio}. Building on this foundation, recent work has extended response modalities to speech, enabling applications in spoken dialogue.
Notably, systems like SpeechGPT \cite{zhang2023speechgpt}, dGSLM \cite{nguyen2023dgslm}, and Freeze-Omni \cite{wang2024freeze} unify speech processing within LLMs, replacing the traditional ASR-TTS pipeline for end-to-end speech interaction.
Additionally, some works consider low latency \cite{defossez2024moshi, veluri2024beyond, fang2024llama, xie2024mini}, dialogue context \cite{lin2024spokenllm} and emotion \cite{xue2024chat, liu2023emotionally}, but they only provide speech responses to users. Recently, as previously mentioned, systems like EmpathyEar \cite{fei2024empathyear} and Empatheia \cite{zhang2025empatheia} have begun providing both voice and visual responses to users. However, our work differs in three key aspects: 1) In this work, the text of the target utterance is given, and we focus on the quality and expressiveness of the synthesized audiovisual content. 2) We directly leverage LLMs to model the input multimodal information, avoiding additional and complex feature extraction or style disentanglement processes. 3) The talking-face animations are not only driven by speech signals but also precisely controlled by facial expression sequences.

\subsection{Talking-face Animations Synthesis}
Significant progress has been made in generating speaking avatar videos from facial images and speech. For example, SadTalker \cite{zhang2023sadtalker} predicts 3D motion coefficients from speech to drive a 3D facial renderer, while Emo \cite{tian2024emo} eliminates the need for 3D models by directly generating videos from speech, achieving smooth transitions and identity consistency. However, methods that rely solely on speech can become unstable when the audio signal is weak. To address this, some approaches incorporate additional guidance. TalkingGaussian \cite{li2024talkinggaussian} uses point-based Gaussian primitives to represent facial motion, enabling smooth deformations driven by speech. Other works employ facial landmarks for more precise control—for instance, AniPortrait \cite{wei2024aniportrait} adopts a two-stage Audio2Lmk and Lmk2Video pipeline, and Echomimic \cite{chen2024echomimic} enhances training strategies to support control via speech, landmarks, or both. Inspired by these studies, we adopt a joint control strategy that infers both speech and facial landmarks from dialogue context using LLM, ensuring better alignment with conversation.

\subsection{Multimodal Neural Codec}
In LLMs, 
discrete tokens are particularly favored in large-scale autoregressive models due to their advantages in multimodal learning and sequential modeling. For example, SpeechTokenizer \cite{zhang2023speechtokenizer} and FACodec \cite{ju2024naturalspeech} use vector quantization to decompose speech into components such as content, timbre, and prosody, while BigCodec \cite{xin2024bigcodec} and Single-Codec \cite{li2024singlecodec} reduce token sequence length through low-bitrate tokenization.
Cosyvoice \cite{du2024cosyvoice} stabilizes sequence modeling by extracting semantically rich speech tokens via an ASR-based \cite{radford2023whisper} self-supervised approach, while Cosyvoice2 \cite{du2024cosyvoice2} improves codebook utilization through Finite Scalar Quantization (FSQ) \cite{mentzer2023fsq}. Following this line of work, we adopt self-supervised learning and FSQ to extract speech tokens from conversational speech.

Similarly, some methods start to tokenize images and videos. For example, \cite{lee2021vitgan} leverage VQGAN \cite{esser2021taming} for high-quality image tokenization and reconstruction, while TiTok \cite{yu2024titok} reduces tokens to just 32 per image. Models like OmniTokenizer \cite{wang2024omnitokenizer} and Qwen2.5-VL \cite{bai2025qwen2vl} extend tokenization to both images and videos. However, directly tokenizing entire talking-face animations introduces redundant information—such as background content—that hinders dialogue context modeling. To better capture facial expressions and enable direct control of reference portraits, we directly tokenize facial keypoint coordinates (landmarks). Moreover, existing methods often produce long token sequences, increasing the difficulty of contextual modeling in LLMs. To address this, we propose \textbf{LmkCodec}, a neural codec that achieves low token rate tokenization while maintaining high reconstruction quality.

\section{Task Definition}
A conversation can be divided into two parts: the conversational history \( \mathcal{H} \), and the current target utterance \( \mathcal{C} \).
Specifically, conversational history \( \mathcal{H} = \{ U_i \}_{i=1}^{N-1} \), where \( U_i \) represents the dialogue utterance in the \( i \)-th turn. Each \( U_i \) consists of multimodal information, including: Speaker \( P_i \), Speech \( U_i^s \), Talking-face animations \( U_i^v \), Text content \( U_i^t \), Emotion category \( U_i^e \). Current target utterance \( \mathcal{C} = \{ U_N \} = \{ P_N, U_N^t \} \), which includes only the speaker and text content.
The model's task is to predict the emotion category \( U_N^{e^{\prime}} \), talking-face animations \( U_N^{v^{\prime}} \), and speech \( U_N^{s^{\prime}} \) of the current utterance based on the \( \mathcal{H} \) and the \( \mathcal{C} \). The task can be formalized as:
\begin{equation}
\resizebox{0.9\columnwidth}{!}{$
\begin{array}{c}
  \{U_N^{e^{\prime}}, U_N^{v^{\prime}}, U_N^{s^{\prime}} \mid \mathcal{H}, \mathcal{C}\}
  \\ =
  \left\{U_N^{e^{\prime}}, U_N^{v^{\prime}}, U_N^{s^{\prime}} \mid \{P_i, U_i^v, U_i^s, U_i^t, U_i^e \}_{i=1}^{N-1} \cup \{ P_N, U_N^t \} \right\}
\end{array}
$}
\end{equation}

The objective is to deeply understand the multimodal dialogue context \( \{\mathcal{H}, \mathcal{C}\} \) and generate \( U_N^{e^{\prime}} \), \( U_N^{s^{\prime}} \), and \( U_N^{v^{\prime}} \) with coherent emotional expression. In addition, the speech \( U_N^{s^{\prime}} \) should not only reflect the target emotion \( U_N^{e^{\prime}} \), but also match the speaking style and timbre of the dialogue context. The animations \( U_N^{v^{\prime}} \) should be aligned with \( U_N^{s^{\prime}} \) in both facial expressions and lip movements, ensuring that speech and visual modalities are synchronized and emotionally coherent.

\section{UniTalker: Methodology}
Fig. \ref{fig:model} illustrates the overall architecture of the UniTalker model. The Multimodal Tokenizer extracts initial representations from the given multimodal context $(\mathcal{H}, \mathcal{C})$. 
(1) In the \textit{multitasking dialogue context modeling} phase, the LLM-based EVSLM model takes serialized context as input. It uses a multi-task learning strategy to jointly learn multimodal cues from the dialogue context and sequentially infer the emotional category, facial expression tokens, and speech tokens for the target utterance.
(2) In the \textit{multimodal responsive content rendering} phase, the decoded emotion category \( U_N^{e^{\prime}} \) and the speech tokens guide the Speech Renderer to produce the empathetic speech \( U_N^{s^{\prime}} \). The empathetic speech, along with the decoded facial expression tokens (facial landmarks) and a reference portrait, feeds into the Talking-face Animations Renderer to generate natural talking-face animations \( U_N^{v^{\prime}} \). Finally, the synthesized speech and talking-face animations are integrated and presented to the user as a coherent multimodal response.

\subsection{Multimodal Dialogue Context Tokenization}
Unlike traditional CSS methods that rely on various pre-trained models to extract complex multi-scale representations \cite{reimers2019sentence, baevski2020wav2vec}, our approach directly tokenizes raw input data, including text, speech, and facial landmarks. And instead of using predefined integer speaker labels, which are unsuitable for speakers not seen during training, we adopt the pre-trained model to extract continuous speaker representations. The implementation details are as follows.

\begin{figure}[t]
\centering
\centerline{
\includegraphics[width=0.95\linewidth]{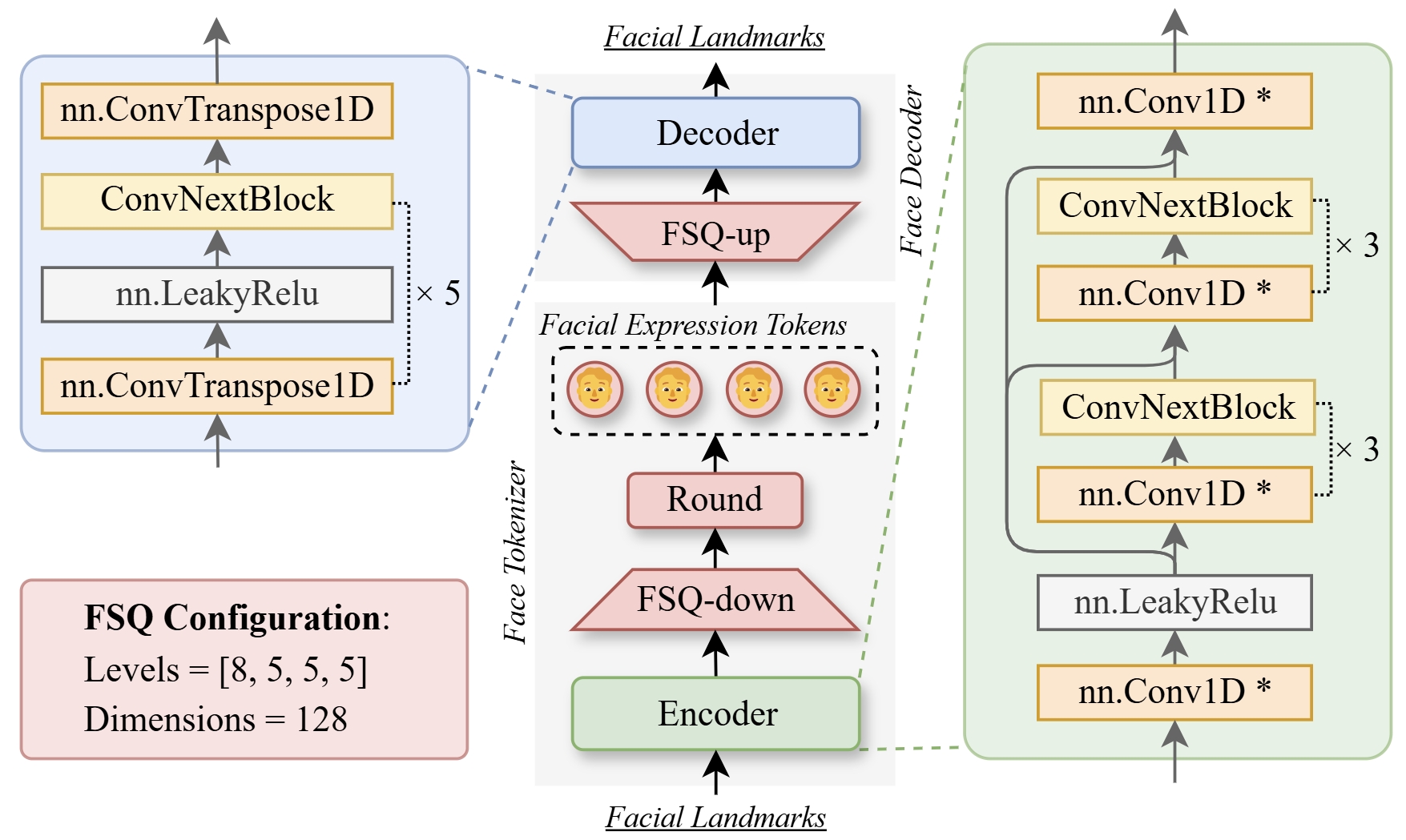}
}
\caption{The overall architecture of LmkCodec, $*$ indicates that left-padding is used.}
\label{fig:codec}
\end{figure}

\subsubsection{\textbf{Text and Emotion Tokenizer}} 
To simplify the processing of textual data and allow the model to learn the text content directly from the context, we use a BPE-based Text Tokenizer \cite{gage1994bpe}. Specifically, the text token sequence corresponding to an utterance \( U_i \) is represented as:
\begin{equation}
  T_{i,<max^{t}}^t = \text{Text-Tokenizer}(U_i^t)
\end{equation}
where \( i \) represents the dialogue turn, \( max^t \) indicates the length of the text token sequence extracted from \( U_i^t \), and $t$ indicates text modality. Additionally, the  UniTalker introduces seven emotion categories \footnote{The emotions include ``Angry'', ``Disgust'', ``Fear'', ``Happy'', ``Neutral'', ``Sadness'', and ``Surprise''.}. To streamline emotion representation, we map the emotions directly using the Text Tokenizer. The emotion category of the utterance \( U_i \) is denoted as $T_{i}^e$.

\subsubsection{\textbf{Speech Tokenizer}}
As previously mentioned, we follow CosyVoice2 \cite{du2024cosyvoice2} and incorporate the FSQ module \cite{mentzer2023fsq} into the encoder of the SenseVoice-Large ASR model \cite{an2024funaudiollm}. Specifically, the intermediate representation produced by the encoder is quantized by the FSQ module. The training process is carried out as an ASR task. The Speech Tokenizer process can be represented as:
\begin{equation}
  T_{i,<max^{s}}^s = \text{Speech-Tokenizer}(U_i^s)
\end{equation}
where \( max^{s} \) denotes the length of the speech token sequence extracted from \( U_i^s \), and $s$ indicates the speech modality.

\subsubsection{\textbf{Face Tokenizer}}
As shown in Fig.~\ref{fig:codec}, we propose a neural codec named \text{LmkCodec} for compressing facial expression sequences. The input is a sequence of facial landmarks \(U_{i, 1 \rightarrow k}^f\), extracted from talking-face animations \(U_i^v\) using Mediapipe \footnote{We use Google's Mediapipe to extract facial landmarks from raw talking-face animations (https://github.com/google-ai-edge/mediapipe).}, where \(k\) is the number of frames. Similar to the Speech Tokenizer, the FSQ module is inserted between the Encoder and Decoder. The Encoder first processes the input to produce hidden features, which are then mapped to a low-rank space using a linear layer (FSQ-down) and quantized into the range \([-K, K]\) by a bounded rounding operation:
\begin{equation}
    T_{i,<max^v}^v = \text{ROUND}(\text{FSQ-down}(\text{Encoder}(U_{i, 1 \rightarrow k}^f)))
\end{equation}

The Encoder, FSQ-down, and ROUND together form the \textbf{Face Tokenizer}. And the quantized tokens \(T_{i,<max^v}^v\) are then mapped back to the original space using another linear layer (FSQ-up), followed by the Decoder to reconstruct the facial landmarks:
\begin{equation}
    U_{i, 1 \rightarrow k}^{f^{\prime}} = \text{Decoder}(\text{FSQ-up}(T_{i,<max^v}^v))
\end{equation}

The FSQ-up and Decoder together form the \textbf{Face Decoder}.  
LmkCodec is trained to minimize the reconstruction loss between the original and predicted landmark sequences using mean squared error (MSE).
The architectures of the Encoder and Decoder are illustrated in Fig.~\ref{fig:codec}.

\subsubsection{\textbf{Speaker Embedding and Special Tokens}}
To better differentiate between speakers during context modeling, a pre-trained voiceprint model \footnote{https://github.com/modelscope/3D-Speaker/tree/main/egs/3dspeaker/sv-cam++} is used to extract speaker embedding \( P_i \) from each utterance's speech \( U_i^s\). For special tokens, they indicate the boundaries of sequences. Specifically, \Circled{B} and \Circled{D} mark the start and end of the entire context sequence. Likewise, \Circled{S} and \Circled{E} denote the start and end of the speech-facial expression token sequence.


\subsection{Multitasking Dialogue Context Modeling}
We propose a large multimodal context modeling module, EVSLM, built upon Qwen2.5-0.5B~\cite{yang2024qwen2.5}. EVSLM is designed to understand utterances and multimodal information within a dialogue context, and sequentially performs multiple subtasks through a step-by-step prediction process.

Recognizing that emotions play a vital role in human communication, EVSLM first infers the emotional state of the speaker and generates the corresponding emotional expression tokens \(T^e_N\) for the target utterance. It then predicts the associated facial expression tokens \(T_{i,<max^v}^v\) and speech tokens \(T_{i,<max^s}^s\).

Unlike previous approaches that predict entire sequences independently~\cite{liu2024gpttalker, wang2023valle}, we introduce a bimodal speech-visual hard alignment strategy that enforces frame-level synchronization between the two modalities. Both speech and facial expression tokens are aligned with the same token rate (e.g., 25 tokens per second), enabling frame-wise modeling. Accordingly, we construct an interleaved input sequence of the form \(\langle T_{N,1}^v, T_{N,1}^s, T_{N,2}^v, T_{N,2}^s, \ldots, T_{N,max^v}^v, \\ T_{N,max^s}^s \rangle\) and perform token-level autoregressive prediction in the same alternating order. This ensures temporal consistency between the synthesized speech and facial expressions, i.e., \(max^s = max^v\), resulting in synchronized lip movements.

During training, we adopt teacher forcing by feeding left-shifted token sequences as inputs and using the original sequences as supervision. The  training objective minimizes the sum of cross-entropy losses for emotion, facial expression tokens, and speech tokens prediction, along with the end-of-sequence token \Circled{D}.
\vspace{-4mm}
\subsection{Multimodal Responsive Content Rendering}
Once the EVSLM context reasoning is complete, UniTalker decodes the predicted tokens for emotion and facial expressions. It then combines them with speech tokens to render empathetic speech and natural talking-face animations.

\vspace{-2mm}
\subsubsection{\textbf{Empathetic Speech Rendering}}
To generate empathetic and expressive conversational speech, we adopt emotion-guided Conditional Flow Matching (CFM)~\cite{mehta2024matcha} for speech rendering. Specifically, we sample Mel spectrograms using a causal convolutional Transformer UNet as the vector field predictor. The model is conditioned on the speaker embedding \(P_N\), the emotion label \(U_N^e\), and a reference Mel spectrogram \(\tilde{X}_1\) which is extracted from the target speaker’s recent speech within the dialogue context, and decodes the predicted speech tokens \(T_{N,<max^s}^s\) into the Mel spectrogram of the target utterance. Finally, the spectrogram is converted into a waveform using the HiFi-GAN vocoder~\cite{kong2020hifi}. To optimize CFM training, we follow CosyVoice2~\cite{du2024cosyvoice2} and employ an Optimal Transport (OT) flow to guide vector field learning. The training objective minimizes the discrepancy between the predicted and ground-truth vector fields.

\subsubsection{\textbf{Natural Talking-face Animations Rendering}}
To achieve natural and expressive talking-face animation synthesis, we adopt Echomimic~\cite{chen2024echomimic}, a state-of-the-art framework, as the backbone of our Talking-face Animations Renderer. Given the target utterance's empathetic speech \(U^s_N\) and facial landmarks \(U^v_{N,<max^v}\), the renderer jointly drives the digital portrait of the target speaker \footnote{It is worth noting that the facial landmarks predicted by EVSLM often do not match the scale of the reference portrait after reconstruction. Therefore, we apply an affine transformation to adjust the landmarks to the target face size before rendering.}. At the core of the renderer is a denoising U-Net, which iteratively refines latent representations while integrating multimodal information through a series of attention mechanisms. Specifically, the Reference-Attention layer incorporates identity features from the reference image to preserve facial consistency. The Audio-Attention layer leverages speech features to drive accurate lip synchronization and expression dynamics, and the Temporal-Attention layer ensures temporal coherence via self-attention over the video sequence.
During training, the model is optimized to maintain both spatial fidelity and temporal consistency by minimizing a combined latent and spatial loss.

\section{Experiments}
In this section, we present the three datasets used for training and describe the multi-stage training strategy in detail. We also introduce three types of baseline models used for comparison, along with the subjective and objective evaluation metrics. 

\subsection{Datasets and Training Strategy}
We utilize three categories of training data to support the UniTalker model:
(1) \textbf{Datasets-1}: Spoken dialogue datasets with contextual information, such as \textit{DailyTalk}~\cite{lee2023dailytalk} and \textit{NCSSD (EN)}~\cite{liu2024gpttalker}, where each dialogue consists of $N$ tuples of \textless{}speaker, text, speech\textgreater{}. The total duration amounts to 113 hours.
(2) \textbf{Datasets-2}: Visual-spoken dialogue datasets with context, such as \textit{MultiDialog}~\cite{park2024multidialog}, where each dialogue contains $N$ tuples of \textless{}speaker, text, speech, video\textgreater{}. We perform emotion recognition on samples labeled as ``Curious to dive deeper'' in the original dataset using Emotion2vec \footnote{\label{emotion2vec}https://huggingface.co/emotion2vec}, and remove 11{,}499 utterances identified as $\text{<unk>}$. The final dataset totals 307 hours.
(3) \textbf{Datasets-3}: Visual-speech datasets in single-utterance settings, including \textit{RAVDESS}~\cite{livingstone2018RAVDESS}, \textit{MEAD}~\cite{wang2020MEAD}, and \textit{CelebV-HQ}~\cite{zhu2022CelebVHQ}, each comprising \textless{}text, speech, video\textgreater{} tuples. The total duration is 89 hours.
For more statistical details, please refer to the Appendix.


We employ a multi-stage training strategy to progressively enhance the multimodal perception and multimodal response capabilities of UniTalker:
(1) \textbf{Stage-1}: We train the LmkCodec using talking-face animations segments from the Datasets-2 and Datasets-3, enabling it to tokenize (via the Face Tokenizer) and reconstruct (via the Face Decoder) facial expression sequences.
(2) \textbf{Stage-2}: We initialize weights of EVSLM and Speech Renderer using pre-trained CosyVoice2 model \footnote{https://huggingface.co/spaces/FunAudioLLM/CosyVoice2-0.5B}, trained on a single-sentence speech dataset of 170,000 hours. Newly introduced modules are randomly initialized.
(3) \textbf{Stage-3}: We fine-tune EVSLM and Speech Renderer with the Datasets-1 to improve their contextual semantic and emotional understanding.
(4) \textbf{Stage-4}: We further fine-tune the \textit{EVSLM} using the Datasets-2 and Datasets-3 to enhance multimodal context understanding capabilities, treating the Datasets-3 as dialogues with a context length of one.
Note that we do not perform additional fine-tuning on the \textit{Talking-face Animations Renderer}, as the official pre-trained weights already demonstrate robust and high-quality talking-face animations rendering capabilities.

\subsection{Baseline Models}
To comprehensively evaluate the performance of UniTalker, we compare it with state-of-the-art models in three areas:
\textbf{(1) \textit{Tokenizing Facial Expression Sequences in Animations}:} We assess LmkCodec's ability to tokenize facial landmarks by comparing it with four baseline models. Currently, to the best of our knowledge, no open-source models are capable of directly tokenizing and reconstructing landmark sequences. Therefore, we use models such as \textit{Taming-VQGAN} \cite{esser2021taming} and \textit{OmniTokenizer} \cite{wang2024omnitokenizer}, which reconstruct images and videos. We extract video frames from the talking-face animations and input them directly into these models for tokenization. After reconstruction, we re-extract the landmarks from each frame. Additionally, we compare UniTalker with two variants of LmkCodec: \textit{LmkCodec-GRVQ} and \textit{LmkCodec-VQ}, which replace the FSQ layer with GRVQ (comprising two groups of four residual VQ layers) \cite{yang2023hificodec} and a single-layer VQ \cite{van2017vq}, respectively.
\textbf{(2) \textit{Synthesizing Speech in Conversational Scenarios}:} We evaluate UniTalker's ability to synthesize speech within dialogue contexts by comparing it with the CSS models that consider dialogue context. These models include the GRU-based context modeling method (\textit{GRU-CSS}) \cite{guo2021conversational}, \textit{M$^2$-CTTS} \cite{xue2023m2ctts}, the MSRGCN-based context modeling method (\textit{MSRGCN-CSS}) \cite{li2022inferring}, \textit{ECSS} \cite{liu2024emotion}, \textit{GPT-Talker} \cite{liu2024gpttalker}, \textit{Empatheia${^\S}$} \cite{zhang2025empatheia}, \textit{EmpathyEar${^\S}$} \cite{fei2024empathyear} and UniTalker-GRVQ (UniTalker uses the LmkCodec-GRVQ). The ${\S}$ denotes the removal of the visual modality from these models.
\textbf{(3) \textit{Generating Talking-face Animations in Conversational Scenarios}:} We evaluate UniTalker's ability to generate talking-face animations within dialogue contexts by comparing it with seven models based on speech-driven reference portrait. These include \textit{AniPortrait} \cite{wei2024aniportrait}, \textit{Hallo} \cite{xu2024hallo}, \textit{Echomimic} \cite{chen2024echomimic}, and \textit{SadTalker} \cite{zhang2023sadtalker}, which drive the speaker's portrait using single speech, and \textit{Empatheia} \cite{zhang2025empatheia} and \textit{EmpathyEar} \cite{fei2024empathyear}, which generate talking-face animations by considering dialogue context. Among them, the models that utilize single speech incorporate speech generated by UniTalker.

\subsection{Metrics}
We use objective and subjective metrics to evaluate each model's performance:

\textbf{\textit{Objective Evaluation Metrics}:}
To measure the quality of generated talking-face animations (format is ``.mp4''), we calculate Fréchet Inception Distance (\textit{FID}) \cite{heusel2017FID}, Peak Signal-to-Noise Ratio (\textit{PSNR}) \cite{sheikh2006PSNR}, Learned Perceptual Image Patch Similarity (\textit{LPIPS}) \cite{zhang2018LPIPS}, and Structural Similarity Index Measure (\textit{SSIM}) \cite{wang2004SSIM} for each frame of the video. These metrics quantify the differences between the generated videos and the real ones. When evaluating LmkCodec, the videos compared are frames rendered solely from the speaker's facial landmarks without any background. Landmark Distance (\textit{LMD}) directly compares the coordinates of the facial landmark sequences. Lip Sync Error Confidence (\textit{LSE-C}) and Lip Sync Error Distance (\textit{LSE-D}) are used to assess the synchronization between the speaker's lip movements and the speech \cite{prajwal2020lip}. For synthesized speech (format is ``.wav'') quality, we compute the speaker similarity (\textit{SIM$_{SPK}$}) \cite{liu2024gpttalker} and the Dynamic Time Warping Distance of pitch distributions (\textit{PDTW}) \cite{liu2024gpttalker} between generated and real speech. Additionally, the emotion recognition models Emotion2vec \footref{emotion2vec} and R1-Omni \cite{zhao2025r1} are used to evaluate the emotional accuracy of the synthesized speech and talking-face animations, measured as \textit{ACC$_{SE}$} and \textit{ACC$_{VE}$}, respectively.

    
\textbf{\textit{Subjective Evaluation Metrics}:}
We recruited 30 trained university students with strong English skills to perform subjective evaluations. The evaluation content includes:
(a) The quality and naturalness of the synthesized speech in a dialog setting (\textit{MOS$_{SN}$}).
(b) The emotional appropriateness and expressiveness of the synthesized speech in a dialog setting (\textit{MOS$_{SE}$}).
(c) The quality and naturalness of the generated talking-face animations in a dialog setting (\textit{MOS$_{VN}$}).
(d) The emotional appropriateness and expressiveness of the generated talking-face animations in a dialog setting (\textit{MOS$_{VE}$}).
(e) The consistency between the speaker's lip movements and the speech in the video (\textit{MOS$_{VC}$}).

\begin{table}[t]
\caption{\label{tab:exp-1} Quantitative comparison between the proposed FSQ-based LmkCodec and baseline models. The resolution is $512 \times 512$. T-rate indicates the number of tokens per frame and V-size denotes the codebook size.}
\vspace{-2mm}
\centering
\resizebox{1\linewidth}{!}{
\begin{tabular}{ccccccc}
\toprule
\multicolumn{1}{c|}{\textbf{Models}} & \multicolumn{1}{c|}{\textbf{T-rate}} & \multicolumn{1}{c|}{\textbf{V-size}} & \textbf{FID $(\downarrow)$} & \textbf{PSNR $(\uparrow)$} & \textbf{LPIPS $(\downarrow)$} & \textbf{LMD $(\downarrow)$} \\ \midrule
\multicolumn{7}{l}{\cellcolor[HTML]{EFEFEF}\textit{The process of reconstruction (data flow): images ~-\textgreater~ tokens ~-\textgreater~ images ~-\textgreater~ landmarks}} \\ \midrule
\multicolumn{1}{c|}{Taming-VQGAN} & \multicolumn{1}{c|}{64 × 64} & \multicolumn{1}{c|}{8192} & 3.513 & 28.696 & \underline{0.485 × 10$^{-2}$} & 0.146 × 10$^{-2}$ \\
\multicolumn{1}{c|}{OmniTokenizer} & \multicolumn{1}{c|}{64 × 64} & \multicolumn{1}{c|}{8192} & \cellcolor[RGB]{245,248,253}\textbf{3.329} & \cellcolor[RGB]{245,248,253}\textbf{30.054} & \cellcolor[RGB]{245,248,253}\textbf{0.457 × 10$^{-2}$} &\cellcolor[RGB]{245,248,253}\textbf{0.115 × 10$^{-2}$}\\\midrule
\multicolumn{7}{l}{\cellcolor[HTML]{EFEFEF}\textit{The process of reconstruction (data flow): landmarks ~-\textgreater~ tokens~-\textgreater~ landmarks}} \\ \midrule
\multicolumn{1}{c|}{LmkCodec-GRVQ} & \multicolumn{1}{c|}{2 × 4} & \multicolumn{1}{c|}{1024} & \underline{3.466} & \underline{29.104} & 0.507 × 10$^{-2}$ & \underline{0.134 × 10$^{-2}$} \\
\multicolumn{1}{c|}{LmkCodec-VQ} & \multicolumn{1}{c|}{1} & \multicolumn{1}{c|}{1024} & 4.673 & 24.596 & 0.572 × 10$^{-2}$ & 0.214 × 10$^{-2}$ \\
\multicolumn{1}{c|}{LmkCodec (ours)} & \multicolumn{1}{c|}{\cellcolor[RGB]{245,248,253}\textbf{1}} & \multicolumn{1}{c|}{\cellcolor[RGB]{245,248,253}\textbf{1000}} & 4.155 & 26.143 & 0.523 × 10$^{-2}$ & 0.185 × 10$^{-2}$ \\ \bottomrule
\end{tabular}
}
\vspace{-4mm}
\end{table}

\section{Results and Discussions}
In this section, we provide a comprehensive evaluation of UniTalker. We first verify the effectiveness of the proposed LmkCodec in tokenizing and reconstructing facial landmarks. Then, we evaluate UniTalker's performance in generating expressive speech and natural talking-face animations. To assess the contribution of each component, we conduct ablation studies. Finally, we present visualization results to intuitively demonstrate the quality of the generated talking-face animations.

\subsection{Reliability Verification of LmkCodec}
In this experiment, we randomly select 200 talking-face animations from the MultiDialog test set for reconstruction and objective evaluation. As shown in rows 2–6 of Table \ref{tab:exp-1}, LmkCodec-GRVQ uses only 8 tokens per frame while achieving performance comparable to the state-of-the-art OmniTokenizer, which directly reconstructs images. Moreover, directly reconstructing facial landmarks removes the need for a separate keypoint extraction step during decoding in the UniTalker Face Decoder, thereby reducing cascade stages. Under the same configuration as LmkCodec-GRVQ, we further evaluate LmkCodec-VQ and LmkCodec with a lower token rate (rows 7 and 8 in Table \ref{tab:exp-1}). Compared to LmkCodec-VQ, the FSQ-based LmkCodec achieves better performance and reduces the codebook size to 1000.

\begin{table}[t]
\caption{\label{tab:exp-2}\textcolor{black}{Subjective (95$\%$ confidence interval) and objective results on expressive speech. $\S $ denotes the removal of visual modal information from the model.}} 
\vspace{-1mm}
\centering
\resizebox{1\linewidth}{!}{
\begin{tabular}{cccccc}
\toprule
\textbf{Models} & \textbf{SIM$_{SPK}$ $(\uparrow)$} & \textbf{PDTW $(\downarrow)$} & \multicolumn{1}{c|}{\textbf{ACC$_{SE}$ $(\uparrow)$}} & \textbf{MOS$_{SN}$ $(\uparrow)$} & \textbf{MOS$_{SE}$ $(\uparrow)$}\\\midrule
\multicolumn{6}{l}{\cellcolor[HTML]{EFEFEF}\textit{Dataset: DailyTalk and NCSSD}} \\\midrule
\multicolumn{1}{c|}{Ground Truth} & N/A & N/A & \multicolumn{1}{c|}{N/A} & \cellcolor[RGB]{250,239,239}{4.518 $_{\pm 0.015}$} & \cellcolor[RGB]{250,239,239}{4.315 $_{\pm 0.015}$} \\
\multicolumn{1}{c|}{GRU-CSS} & 0.754 & 66.423 & \multicolumn{1}{c|}{0.536} & 3.412 $_{\pm 0.024}$ & 3.436 $_{\pm 0.021}$ \\
\multicolumn{1}{c|}{M$^2$-CTTS} & 0.759 & 64.124 & \multicolumn{1}{c|}{0.624} & 3.623 $_{\pm 0.018}$ & 3.597 $_{\pm 0.016}$ \\
\multicolumn{1}{c|}{MSRGCN-CSS} & 0.762 & 63.217 & \multicolumn{1}{c|}{0.633} & 3.683 $_{\pm 0.034}$ & 3.651 $_{\pm 0.024}$ \\
\multicolumn{1}{c|}{ECSS} & 0.764 & 59.578 & \multicolumn{1}{c|}{0.651} & 3.764 $_{\pm 0.011}$ & 3.768 $_{\pm 0.011}$ \\
\multicolumn{1}{c|}{GPT-Talker} & \underline{0.879} & \underline{44.156} & \multicolumn{1}{c|}{0.674} & 3.917 $_{\pm 0.025}$ & 3.907 $_{\pm 0.033}$ \\
\multicolumn{1}{c|}{EmpathyEar$^\S $} & 0.864 & 47.274 & \multicolumn{1}{c|}{0.701} & 3.921 $_{\pm 0.031}$ & 3.934 $_{\pm 0.020}$ \\
\multicolumn{1}{c|}{Empatheia$^\S $} & 0.868 & 46.355 & \multicolumn{1}{c|}{\underline{0.713}} & \underline{4.036 $_{\pm 0.016}$} & \underline{3.955 $_{\pm 0.019}$} \\
\multicolumn{1}{c|}{UniTalker$^\S $} & \cellcolor[RGB]{245,248,253}\textbf{0.903} & \cellcolor[RGB]{245,248,253}\textbf{41.043} & \multicolumn{1}{c|}{\cellcolor[RGB]{245,248,253}\textbf{0.743}} & \cellcolor[RGB]{245,248,253}\textbf{4.142 $_{\pm 0.028}$} & \cellcolor[RGB]{245,248,253}\textbf{4.116 $_{\pm 0.027}$} \\\midrule
\multicolumn{6}{l}{\cellcolor[HTML]{EFEFEF}\textit{Dataset: MultiDialog}} \\\midrule
\multicolumn{1}{c|}{Ground Truth} & N/A & N/A & \multicolumn{1}{c|}{N/A} & \cellcolor[RGB]{250,239,239}{4.524 $_{\pm 0.032}$} & \cellcolor[RGB]{250,239,239}{4.315 $_{\pm 0.012}$} \\
\multicolumn{1}{c|}{GRU-CSS} & 0.732 & 68.211 & \multicolumn{1}{c|}{0.552} & 3.401 $_{\pm 0.006}$ & 3.447 $_{\pm 0.015}$ \\
\multicolumn{1}{c|}{M$^2$-CTTS} & 0.748 & 67.024 & \multicolumn{1}{c|}{0.627} & 3.618 $_{\pm 0.017}$ & 3.601 $_{\pm 0.026}$ \\
\multicolumn{1}{c|}{MSRGCN-CSS} & 0.752 & 64.247 & \multicolumn{1}{c|}{0.631} & 3.637 $_{\pm 0.032}$ & 3.647 $_{\pm 0.019}$ \\
\multicolumn{1}{c|}{ECSS} & 0.758 & 58.158 & \multicolumn{1}{c|}{0.656} & 3.755 $_{\pm 0.024}$ & 3.762 $_{\pm 0.032}$ \\
\multicolumn{1}{c|}{GPT-Talker} & \underline{0.871} & \underline{47.271} & \multicolumn{1}{c|}{0.681} & 3.912 $_{\pm 0.018}$ & 3.913 $_{\pm 0.022}$ \\ 
\multicolumn{1}{c|}{EmpathyEar$^\S $} & 0.823 & 50.493 & \multicolumn{1}{c|}{0.711} & 3.914 $_{\pm 0.035}$ & 3.942 $_{\pm 0.016}$ \\
\multicolumn{1}{c|}{Empatheia$^\S $} & 0.834 & 48.017 & \multicolumn{1}{c|}{\underline{0.718}} & \underline{3.932 $_{\pm 0.013}$} & \underline{3.953 $_{\pm 0.028}$} \\
\multicolumn{1}{c|}{UniTalker$^\S $} & \cellcolor[RGB]{245,248,253}\textbf{0.891} & \cellcolor[RGB]{245,248,253}\textbf{42.241} & \multicolumn{1}{c|}{\cellcolor[RGB]{245,248,253}\textbf{0.739}} & \cellcolor[RGB]{245,248,253}\textbf{4.121 $_{\pm 0.019}$} & \cellcolor[RGB]{245,248,253}\textbf{4.076 $_{\pm 0.013}$} \\ \midrule
\multicolumn{1}{c|}{EmpathyEar} & 0.827 & 49.848 & \multicolumn{1}{c|}{\underline{0.726}} & 3.938 $_{\pm 0.017}$ & 3.978 $_{\pm 0.024}$ \\
\multicolumn{1}{c|}{Empatheia} & \underline{0.833} & \underline{47.215} & \multicolumn{1}{c|}{0.722} & \underline{4.015 $_{\pm 0.023}$} & \underline{3.992 $_{\pm 0.011}$} \\
\multicolumn{1}{c|}{UniTalker-GRVQ} & 0.773 & 80.348 & \multicolumn{1}{c|}{0.564} & 3.614 $_{\pm 0.013}$ & 3.536 $_{\pm 0.031}$ \\
\multicolumn{1}{c|}{UniTalker} & \cellcolor[RGB]{245,248,253}\textbf{0.902} & \cellcolor[RGB]{245,248,253}\textbf{42.014} & \multicolumn{1}{c|}{\cellcolor[RGB]{245,248,253}\textbf{0.743}} & \cellcolor[RGB]{245,248,253}\textbf{4.128 $_{\pm 0.021}$} & \cellcolor[RGB]{245,248,253}\textbf{4.103 $_{\pm 0.018}$} \\\bottomrule
\end{tabular}
}
\vspace{-4mm}
\end{table}

\vspace{-2mm}
\subsection{Synthesized Speech Evaluation}
\label{sec-exp-2}
Table \ref{tab:exp-2} presents the objective and subjective experimental results comparing UniTalker$^\S$ with state-of-the-art conversational speech synthesis models across three datasets. The symbol $\S$ indicates the removal of visual information from the corresponding model's contextual modeling process. Specifically, rows 2–11 show the first experiment evaluated on DailyTalk and NCSSD, using 50 dialogue samples from the test sets, respectively. Rows 12–21 show the second experiment evaluated on MultiDialog, using 100 dialogue samples from the test set. Rows 22–25 show the third experiment evaluated on MultiDialog, using 100 dialogue samples from the test set with visual information.

For the 1st and 2nd experiments, comparing SIM$_{SPK}$, UniTalker$^\S$ outperforms GPT-Talker by 0.024 and 0.020, respectively. This indicates that UniTalker$^\S$ better preserves speaker timbre and achieves higher speaker consistency. Additionally, the comparison of PDTW, which relates to speech expressiveness, demonstrates that UniTalker$^\S$ produces speech with superior stylistic expressiveness. MOS$_{SN}$ scores of UniTalker$^\S$ surpass those of Empatheia$^\S$ by 0.106 and 0.189, suggesting UniTalker$^\S$ generates more natural speech. And comparisons based on ACC$_{SE}$ and MOS$_{SE}$ reveal that UniTalker$^\S$ has more accurate emotional understanding and rendering abilities. For the 3rd experiment, UniTalker achieves the best performance in both objective and subjective evaluations when multimodal dialogue contexts containing visual modality are considered. Particularly, UniTalker outperforms the second-best model by 0.017 and 0.111 in terms of emotional expression metrics ACC$_{SE}$ and MOS$_{SE}$, respectively. Furthermore, Table \ref{tab:exp-1} shows that UniTalker-GRVQ with GRVQ structure provides better reconstruction quality compared to a single-layer FSQ. However, results from Table \ref{tab:exp-2} indicate that using 8 tokens per frame increases the complexity of contextual modeling and inference in EVSLM, leading to reduced performance. Therefore, UniTalker effectively balances contextual inference performance with the reconstruction quality provided by LmkCodec. Overall, these experimental results confirm that UniTalker achieves superior speech synthesis quality and enhanced expressiveness by effectively perceiving multimodal context in dialogue scenarios.

\begin{table}[t]
\caption{\label{tab:exp-3}\textcolor{black}{Subjective (95$\%$ confidence interval) and objective results on responsive talking-face animations.}} 
\centering
\resizebox{1\linewidth}{!}{
\begin{tabular}{cccccc}
\toprule
\multicolumn{1}{c|}{\textbf{Models}} & \multicolumn{1}{c}{\textbf{FID $(\downarrow)$}} & \multicolumn{1}{c}{\textbf{PSNR$(\uparrow)$}} & \multicolumn{1}{c}{\textbf{LPIPS $(\downarrow)$}} & \multicolumn{1}{c}{\textbf{SSIM$(\uparrow)$}} & \multicolumn{1}{c}{\textbf{ACC$_{VE}$ $(\uparrow)$}} \\ \midrule
\multicolumn{1}{l|}{Ground Truth} & \cellcolor[RGB]{250,239,239}{0.000} & \cellcolor[RGB]{250,239,239}{$\infty$} & \cellcolor[RGB]{250,239,239}{0.000} & \cellcolor[RGB]{250,239,239}{1.000} & N/A \\ \midrule
\multicolumn{6}{l}{\cellcolor[HTML]{EFEFEF}\textit{Data Format for Model Input: single sentence (speech from UniTalker)}} \\ \midrule
\multicolumn{1}{c|}{SadTalker} & \multicolumn{1}{c}{44.472} & \multicolumn{1}{c}{16.077} & \multicolumn{1}{c}{0.346} & \multicolumn{1}{c}{0.693} & \multicolumn{1}{c}{0.735} \\
\multicolumn{1}{c|}{AniPortrait} & \multicolumn{1}{c}{29.081} & \multicolumn{1}{c}{18.361} & \multicolumn{1}{c}{0.218} & \multicolumn{1}{c}{0.712} & \multicolumn{1}{c}{0.623} \\
\multicolumn{1}{c|}{Hallo} & \multicolumn{1}{c}{27.116} & \multicolumn{1}{c}{\cellcolor[RGB]{245,248,253}\textbf{19.614}} & \multicolumn{1}{c}{0.184} & \multicolumn{1}{c}{\underline{0.737}} & \multicolumn{1}{c}{0.801} \\
\multicolumn{1}{c|}{EchoMimic} & \multicolumn{1}{c}{\underline{24.328}} & \multicolumn{1}{c}{18.684} & \multicolumn{1}{c}{\underline{0.210}} & \multicolumn{1}{c}{0.724} & \multicolumn{1}{c}{\underline{0.807}} \\
 \midrule
\multicolumn{6}{l}{\cellcolor[HTML]{EFEFEF}\textit{Data Format for Model Input: dialogue}} \\ \midrule
\multicolumn{1}{c|}{EmpathyEar} & \multicolumn{1}{c}{74.626} & \multicolumn{1}{c}{13.780} & \multicolumn{1}{c}{0.429} & \multicolumn{1}{c}{0.651} & \multicolumn{1}{c}{0.769} \\
\multicolumn{1}{c|}{Empatheia} & \multicolumn{1}{c}{41.287} & \multicolumn{1}{c}{17.908} & \multicolumn{1}{c}{0.292} & \multicolumn{1}{c}{0.723} & \multicolumn{1}{c}{0.782} \\
\multicolumn{1}{c|}{UniTalker-GRVQ} & \multicolumn{1}{c}{58.632} & \multicolumn{1}{c}{15.735} & \multicolumn{1}{c}{0.358} & \multicolumn{1}{c}{0.664} & \multicolumn{1}{c}{0.506} \\
\multicolumn{1}{c|}{UniTalker} & \multicolumn{1}{c}{\cellcolor[RGB]{245,248,253}\textbf{24.214}} & \multicolumn{1}{c}{\underline{19.513}} & \multicolumn{1}{c}{\cellcolor[RGB]{245,248,253}\textbf{0.204}} & \multicolumn{1}{c}{\cellcolor[RGB]{245,248,253}\textbf{0.741}} & \multicolumn{1}{c}{\cellcolor[RGB]{245,248,253}\textbf{0.813}} \\ \midrule
\multicolumn{1}{c|}{\textbf{Models}} & \multicolumn{1}{c}{\textbf{LSE-C$(\uparrow)$}} & \multicolumn{1}{c|}{\textbf{LSE-D$(\downarrow)$}} & \multicolumn{1}{c}{\textbf{MOS$_{VC}$$(\uparrow)$}} & \multicolumn{1}{c}{\textbf{MOS$_{VN}$$(\uparrow)$}} & \multicolumn{1}{c}{\textbf{MOS$_{VE}$$(\uparrow)$}} \\ \midrule
\multicolumn{1}{c|}{Ground Truth} & \multicolumn{1}{c}{\cellcolor[RGB]{250,239,239}{7.114}} & \multicolumn{1}{c|}{\cellcolor[RGB]{250,239,239}{7.551}} & \multicolumn{1}{c}{\cellcolor[RGB]{250,239,239}{4.346 $_{\pm 0.016}$}} & \multicolumn{1}{c}{\cellcolor[RGB]{250,239,239}{4.556 $_{\pm 0.023}$}}  & \multicolumn{1}{c}{\cellcolor[RGB]{250,239,239}{4.418 $_{\pm 0.012}$}} \\ \midrule
\multicolumn{6}{l}{\cellcolor[HTML]{EFEFEF}\textit{Data Format for Model Input: single sentence (speech from UniTalker)}} \\ \midrule
\multicolumn{1}{c|}{SadTalker} & \multicolumn{1}{c}{6.263} & \multicolumn{1}{c|}{8.342} & \multicolumn{1}{c}{4.096 $_{\pm 0.031}$} & \multicolumn{1}{c}{3.986 $_{\pm 0.025}$} & \multicolumn{1}{c}{3.958 $_{\pm 0.014}$} \\ 
\multicolumn{1}{c|}{AniPortrait} & \multicolumn{1}{c}{3.427} & \multicolumn{1}{c|}{10.769} & \multicolumn{1}{c}{3.574 $_{\pm 0.027}$} & \multicolumn{1}{c}{3.715 $_{\pm 0.018}$} & \multicolumn{1}{c}{3.828 $_{\pm 0.021}$} \\
\multicolumn{1}{c|}{Hallo} & \multicolumn{1}{c}{\cellcolor[RGB]{245,248,253}\textbf{6.412}} & \multicolumn{1}{c|}{\cellcolor[RGB]{245,248,253}\textbf{8.014}} & \multicolumn{1}{c}{\underline{4.135 $_{\pm 0.017}$}} & \multicolumn{1}{c}{4.031 $_{\pm 0.034}$} & \multicolumn{1}{c}{4.163 $_{\pm 0.014}$} \\
\multicolumn{1}{c|}{EchoMimic} & \multicolumn{1}{c}{6.295} & \multicolumn{1}{c|}{8.293} & \multicolumn{1}{c}{4.117 $_{\pm 0.016}$} & \multicolumn{1}{c}{\underline{4.128 $_{\pm 0.009}$}} & \multicolumn{1}{c}{\underline{4.205 $_{\pm 0.036}$}} \\ 
\midrule
\multicolumn{6}{l}{\cellcolor[HTML]{EFEFEF}\textit{Data Format for Model Input: dialogue}} \\ \midrule
\multicolumn{1}{c|}{EmpathyEar} & \multicolumn{1}{c}{5.643} & \multicolumn{1}{c|}{8.874} & \multicolumn{1}{c}{3.718 $_{\pm 0.026}$} & \multicolumn{1}{c}{3.628 $_{\pm 0.012}$} & \multicolumn{1}{c}{3.989 $_{\pm 0.022}$} \\
\multicolumn{1}{c|}{Empatheia} & \multicolumn{1}{c}{6.042} & \multicolumn{1}{c|}{8.574} & \multicolumn{1}{c}{3.924 $_{\pm 0.011}$} & \multicolumn{1}{c}{3.819 $_{\pm 0.031}$} & \multicolumn{1}{c}{4.088 $_{\pm 0.013}$} \\
\multicolumn{1}{c|}{UniTalker-GRVQ} & \multicolumn{1}{c}{2.153} & \multicolumn{1}{c|}{14.857} & \multicolumn{1}{c}{3.126 $_{\pm 0.022}$} & \multicolumn{1}{c}{3.218 $_{\pm 0.014}$} & \multicolumn{1}{c}{3.278 $_{\pm 0.031}$} \\ 
\multicolumn{1}{c|}{UniTalker} & \multicolumn{1}{c}{\underline{6.386}} & \multicolumn{1}{c|}{\underline{8.189}} & \multicolumn{1}{c}{\cellcolor[RGB]{245,248,253}\textbf{4.148 $_{\pm 0.015}$}} & \multicolumn{1}{c}{\cellcolor[RGB]{245,248,253}\textbf{4.254 $_{\pm 0.029}$}} & \multicolumn{1}{c}{\cellcolor[RGB]{245,248,253}\textbf{4.323 $_{\pm 0.017}$}} \\ \bottomrule
\end{tabular}
}
\end{table}

\begin{table*}[t]
\caption{\label{tab:exp-4}\textcolor{black}{Subjective (95\% confidence interval) and objective results of the ablation study on UniTalker. E-guided means adding emotional guidance to the Speech Renderer module. Alignment refers to the bimodal speech-visual hard alignment decoding strategy; otherwise, the two token sequences would be predicted independently. D-history means the conversational history of the input model.}} 
\vspace{-2mm}
\centering
\resizebox{1\linewidth}{!}{
\begin{tabular}{l|ccc|cc|ccc|ccc}
\toprule
\multicolumn{1}{c|}{\textbf{Models}} & \textbf{SIM$_{SPK}(\uparrow)$} & \textbf{PDTW$(\downarrow)$} & \textbf{ACC$_{SE}(\uparrow)$} & \textbf{MOS$_{SN}(\uparrow)$} & \textbf{MOS$_{SE}(\uparrow)$} & \textbf{FID$(\downarrow)$} & \textbf{ACC$_{VE}(\uparrow)$} & \textbf{LSE-C$(\uparrow)$} & \textbf{MOS$_{VC}(\uparrow)$} & \textbf{MOS$_{VN}(\uparrow)$} & \textbf{MOS$_{VE}(\uparrow)$} \\ \midrule
\multicolumn{1}{c|}{UniTalker} & \cellcolor[RGB]{245,248,253}\textbf{0.902} & \cellcolor[RGB]{245,248,253}\textbf{42.014} & \cellcolor[RGB]{245,248,253}\textbf{0.743} & \cellcolor[RGB]{245,248,253}\textbf{4.128$_{\pm 0.021}$} & \cellcolor[RGB]{245,248,253}\textbf{4.103$_{\pm 0.018}$} & \cellcolor[RGB]{245,248,253}\textbf{24.214} & \cellcolor[RGB]{245,248,253}\textbf{0.813} & \cellcolor[RGB]{245,248,253}\textbf{6.386} & \cellcolor[RGB]{245,248,253}\textbf{4.135$_{\pm 0.015}$} & \cellcolor[RGB]{245,248,253}\textbf{4.154$_{\pm 0.029}$} & \cellcolor[RGB]{245,248,253}\textbf{4.213$_{\pm 0.017}$}\\ \midrule
~{-w/o} E-guided & 0.894 & 50.367 & 0.682 & 3.925$_{\pm 0.026}$ & 3.916$_{\pm 0.015}$ & 26.376 & 0.753 & 6.284 & 4.015$_{\pm 0.037}$ & 4.026$_{\pm 0.016}$ & 4.008$_{\pm 0.021}$ \\
~{-w/o} $\mathcal{L}^{emotion}$ & 0.871 & 54.964 & 0.642 & 3.946$_{\pm 0.031}$ & 3.894$_{\pm 0.016}$ & 28.657 & 0.748 & 6.103 & 3.987$_{\pm 0.036}$ & 3.994$_{\pm 0.018}$ & 3.975$_{\pm 0.019}$ \\
~{-w/o} $\mathcal{L}^{face}$ & 0.826 & 65.845 & 0.597 & 3.672$_{\pm 0.043}$ & 3.608$_{\pm 0.017}$ & 60.486 & 0.607 & 5.163 & 3.707$_{\pm 0.018}$ & 3.709$_{\pm 0.043}$ & 3.768$_{\pm 0.016}$ \\
~{-w/o} Stage-2 & 0.751 & 66.238 & 0.601 & 3.601$_{\pm 0.026}$ & 3.643$_{\pm 0.029}$ & 50.781 & 0.657 & 5.667 & 3.735$_{\pm 0.028}$ & 3.807$_{\pm 0.013}$ & 3.792$_{\pm 0.037}$ \\
~{-w/o} Stage-3 & 0.886 & 45.238 & 0.733 & 4.015$_{\pm 0.017}$ & \underline{4.083$_{\pm 0.024}$} & 24.961 & 0.796 & 6.194 & 4.093$_{\pm 0.016}$ & \underline{4.116$_{\pm 0.009}$} & 4.103$_{\pm 0.028}$ \\
~{-w/o} Alignment & 0.781 & 72.684 & 0.613 & 3.693$_{\pm 0.022}$ & 3.683$_{\pm 0.032}$ & 59.127 & 0.572 & 4.452 & 3.567$_{\pm 0.019}$ & 3.659$_{\pm 0.021}$ & 3.612$_{\pm 0.016}$ \\
~{-w/o} D-history & 0.853 & 60.123 & 0.587 & 3.737$_{\pm 0.014}$ & 3.642$_{\pm 0.026}$ & 53.127 & 0.648 & 6.127 & 3.867$_{\pm 0.034}$ & 3.827$_{\pm 0.013}$ & 3.646$_{\pm 0.022}$ \\
~{-w/o} U$^t_{1 \rightarrow N-1}$ & 0.892 & 44.651 & \underline{0.738} & \underline{4.103$_{\pm 0.011}$} & 4.006$_{\pm 0.028}$ & 25.168 & \underline{0.801} & \underline{6.343} & \underline{4.112$_{\pm 0.013}$} & 4.032$_{\pm 0.025}$ & \underline{4.167$_{\pm 0.023}$} \\
~{-w/o} U$^s_{1 \rightarrow N-1}$ & 0.862 & 48.367 & 0.712 & 3.989$_{\pm 0.037}$ & 3.972$_{\pm 0.015}$ & 26.367 & 0.785 & 6.313 & 4.098$_{\pm 0.022}$ & 3.976$_{\pm 0.018}$ & 3.986$_{\pm 0.037}$ \\
~{-w/o} U$^v_{1 \rightarrow N-1}$ & 0.889 & \underline{43.194} & 0.732 & 4.068$_{\pm 0.016}$ & 4.027$_{\pm 0.041}$ & 28.319 & 0.762 & 6.178 & 3.924$_{\pm 0.014}$ & 3.984$_{\pm 0.031}$ & 3.994$_{\pm 0.035}$ \\
~{-w/o} U$^e_{1 \rightarrow N-1}$ & \underline{0.896} & 44.128 & 0.718 & 4.038$_{\pm 0.019}$ & 4.037$_{\pm 0.006}$ & 25.367 & 0.768 & 6.245 & 4.035$_{\pm 0.016}$ & 4.086$_{\pm 0.006}$ & 4.102$_{\pm 0.013}$ \\
~{-w/o} P$_{1 \rightarrow N-1}$ & 0.871 & 45.328 & 0.726 & 4.062$_{\pm 0.023}$ & 4.052$_{\pm 0.011}$ & \underline{24.869} & 0.798 & 6.271 & 4.007$_{\pm 0.022}$ & 4.052$_{\pm 0.014}$ & 4.156$_{\pm 0.026}$ \\ \bottomrule
\end{tabular}
}
\end{table*}

\begin{figure*}[t]
\centering
\centerline{
\includegraphics[width=1\linewidth]{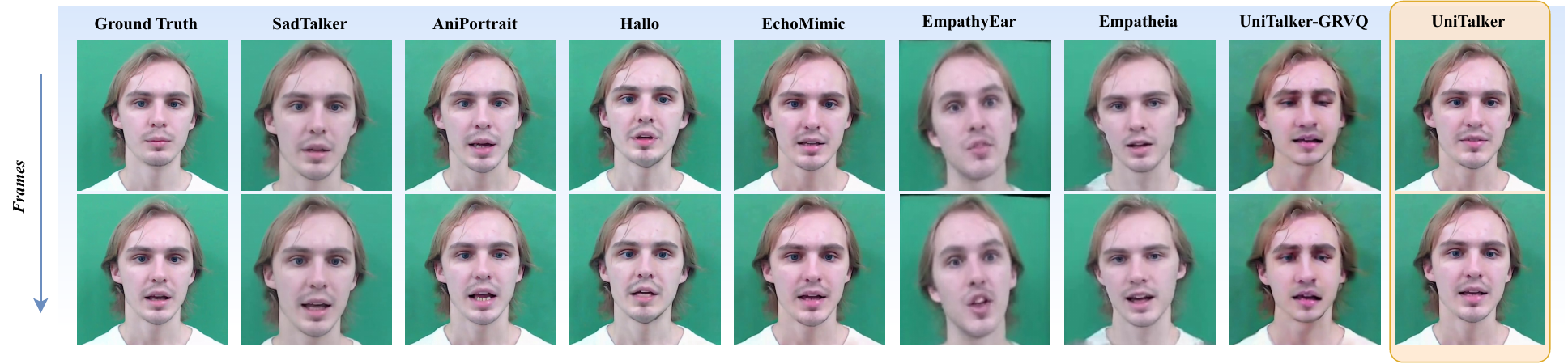}
}
\vspace{-3mm}
\caption{Visualization of talking-face animations generated by UniTalker and baseline models on the MultiDialog dataset.}
\label{fig:visiual}
\vspace{-2mm}
\end{figure*}

\subsection{Generated Talking-face Animation Evaluation}
We comprehensively evaluate UniTalker's capability to generate talking-face animations using the MultiDialog dataset. Specifically, we select 100 dialogue samples from the test set for evaluation. It is important to note that the lengths of the generated and real animations are usually not the same. Therefore, we extract 50 frames at corresponding phoneme-aligned positions from each dialogue sample to compute objective metrics. Table \ref{tab:exp-3} presents the objective and subjective evaluation results. As shown in rows 1–12, UniTalker achieves the best performance in terms of FID, LPIPS, and SSIM, with scores of 24.214, 0.204, and 0.741, respectively. It also secures the second-best PSNR score, trailing speech-driven Hallo by only 0.101. These results demonstrate that UniTalker produces high-quality animations with consistent facial identity and background continuity during both landmark-driven and speech-driven processes. Moreover, UniTalker achieves the highest ACC$_{VE}$ score of 0.813, indicating its superior accuracy in facial emotional expressions. Rows 13–24 show that UniTalker achieves the second-best results in LSE-C and LSE-D, and the best MOS$_{VC}$, highlighting its effective synchronization between synthesized speech and lip movements. Additionally, UniTalker achieves the highest MOS$_{VN}$ and MOS$_{VE}$ scores, outperforming EchoMimic by 0.126 and 0.118, respectively. These results further confirm that UniTalker generates talking-face animations that are notably more natural and emotionally expressive. Overall, UniTalker significantly outperforms multimodal context-aware models, including EmpathyEar and Empatheia, across all evaluated metrics. Moreover, consistent with the analysis presented in subsection \ref{sec-exp-2}, longer facial expression token sequences substantially increase the prediction difficulty for EVSLM, leading to considerably lower performance compared to UniTalker.

\vspace{-5mm}
\subsection{Ablation Results}
We thoroughly investigate the effects of individual modules, training steps, and multimodal conversational history on the overall performance of UniTalker. Specifically, we select 200 dialogue samples from Multidialog for subjective and objective evaluations. The results are shown in Table \ref{tab:exp-4}. As indicated in rows 3 (``-w/o E-guided'') and 4 (``-w/o $\mathcal{L}^{emotion}$''), removing emotional guidance from the Speech Renderer or emotion label constraints from EVSLM significantly reduces UniTalker’s ability to understand emotions. Row 5 (``-w/o $\mathcal{L}^{face}$'') shows that the face loss is essential for EVSLM to effectively learn sequences of facial expressions. From rows 6 and 7, partially initializing UniTalker with the pretrained CosyVoice2 weights improves training stability, while fine-tuning with the CSS dataset further enhances its understanding of contextual semantics. Row 8 demonstrates the importance of alternating and aligning the speech and visual tokens. Rows 9-14 show that conversational history is essential for UniTalker to accurately render the current utterance. Moreover, multimodal dialogue information influences the target utterance differently in terms of semantics, emotion, facial expression, and speaker identity.

\subsection{Visualization Results}
As shown in Fig. \ref{fig:visiual}, we conduct a visualization experiment to present the quality of the generated talking-face animations in a more intuitive manner. Specifically, we use the same speech to drive the reference portrait. From all the generated animations, we extract frames at the same phoneme positions for visual comparison. The results show that UniTalker outperforms other baseline models in terms of visual clarity, facial detail preservation, lip synchronization, head position, and overall naturalness. Moreover, the outputs of UniTalker are visually very close to the Ground Truth.

\section{Conclusion}
In this work, we extend the traditional Conversational Speech Synthesis (CSS) task, which is limited in contextual awareness and generates only unimodal responses, into a more expressive Conversational Speech-Visual Synthesis (CSVS) task. And we propose \textit{UniTalker}, a unified model that seamlessly integrates multimodal context understanding and content rendering, enabling the generation of audiovisual responses with emotional consistency. To further enhance performance, we design a dedicated neural landmark codec to tokenize and reconstruct facial expression sequences. Moreover, a bimodal speech-visual hard alignment decoding strategy is introduced to ensure temporal and content alignment, while emotion-guided rendering maintains emotional coherence between the synthesized speech and talking-face animations.

\clearpage

\section{Acknowledgments}
The research by Rui Liu was funded by the Young Scientists Fund (No.~62206136), the General Program (No.~62476146) of the National Natural Science Foundation of China,  the Young Elite Scientists Sponsorship Program by CAST (2024QNRC001), the Outstanding Youth Project of Inner Mongolia Natural Science Foundation (2025JQ011), Key R\&D and Achievement Transformation Program of Inner Mongolia Autonomous Region (2025YFHH0014) and the Central Government Fund for Promoting Local Scientific and Technological Development (2025ZY0143).
The research by Yifan Hu was funded by the Research and
Innovation Projects for Graduate Students in Inner Mongolia Autonomous Region.
The work by Haizhou Li was supported by the Shenzhen Science and Technology Program (Shenzhen Key Laboratory, Grant No.~ZDSYS20230626091302006), the Shenzhen Science and Technology Research Fund (Fundamental Research Key Project, Grant No.~JCYJ20220818103001002), and the Program for Guangdong Introducing Innovative and Entrepreneurial Teams, Grant No.~2023ZT\\10X044.

\bibliographystyle{ACM-Reference-Format}
\bibliography{sample-base}

\appendix
\section*{Technical Appendix}
In this Appendix, we provide additional information about UniTalker for reference, including ethical considerations, additional experimental details, supplementary experimental results, limitations, and future Work.

\section{Ethic Considerations}
UniTalker supports zero-shot generation of speech and talking-face animations, making it easier to synthesize personalized content. When used appropriately and legally, this technology can benefit a wide range of applications such as film, gaming, podcasting, and other content services, thereby enhancing human experiences. However, zero-shot CSVS also carries the risk of misuse in deepfake-related scenarios, such as voice spoofing and visual deception. To address these concerns, we plan to incorporate watermarks into the synthesized speech and videos, allowing the public to easily identify whether the content is artificially generated. Our project license will also include usage restrictions to prevent the misuse of the model.

\section{More Details of Experiments}
In this section, we provide additional details of the experiments, including the experimental setup, datasets, baseline models, and subjective evaluation.

\subsection{Experimental Setup}
As shown in Table~\ref{tab:appendix-1}, we present a subset of the key module parameters used in UniTalker. Specifically, we adopt Repetition-Aware Sampling for the EVSLM component, following VALL-E 2 \cite{chen2024valle2}. For more detailed model configurations and parameters, please refer to our open-source code on GitHub. The entire training process of UniTalker is conducted on four NVIDIA A100 GPUs. For all three types of datasets, we follow a 7:2:1 split for training, validation, and testing sets, respectively. During training, the context length for UniTalker is set between 1 and 3. During inference, all compared CSS and CSVS models are evaluated over 3 dialogue turns.

\begin{table}[t]
\caption{\label{tab:appendix-1} Partial hyperparameters of the UniTalker.}
\centering
\resizebox{0.8\linewidth}{!}{
\begin{tabular}{@{}c|c|c@{}}
\toprule
\multirow{11}{*}{\textbf{EVSLM}} & speech\_sample\_rate & 24000 \\
 & speech\_token\_rate & 25 \\
 & face\_token\_rate & 25 \\
 & spk\_embed\_dim & 192 \\
 & llm\_input\_size & 896 \\
 & llm\_output\_size & 896 \\
 & text\_vocab\_size & 151936 \\
 & speech\_vocab\_size & 6561 \\
 & face\_vocab\_size & 1000 \\
 & top\_k & 25 \\
 & top\_p & 0.8 \\ \midrule
\multirow{4}{*}{\textbf{LmkCodec}} & Input\_type & Landmarks \\
 & input\_len\_per\_frame & 190 \\
 & latent\_dim\_per\_frame & 128 \\
 & fsq\_levels & {[}8,5,5,5{]} \\ \midrule
\multirow{4}{*}{\textbf{\shortstack{Speech \\ Renderer}}} & input\_size & 512 \\
 & output\_size & 80 \\
 & output\_type & Mel \\
 & token\_mel\_ratio & 2 \\ \midrule
\multirow{3}{*}{\textbf{\shortstack{Talking-face\\ Animations \\ Renderer}}} & image\_resolution & 512 × 512 \\
 & motion\_module\_resolutions & {[}1,2,4,8{]} \\
 & motion\_module\_type & Vanilla \\ \bottomrule
\end{tabular}
}
\end{table}

\subsection{Datasets}
During the training of UniTalker, we use three types of datasets. To equip UniTalker with emotion understanding capabilities, we adopt seven emotion categories. The emotion label distribution in the MultiDialog dataset is as follows: ``Angry'': 928, ``Disgust'': 1,439, ``Fear'': 923, ``Happy'': 54,447, ``Neutral'': 80,892, ``Sadness'': 7,419, and ``Surprise'': 30,307. For the distribution of emotion labels in the other datasets, please refer to the original dataset sources.

\subsection{Baselines}
\subsubsection{\textbf{First category of baseline models (Tokenizing Facial Expression Sequences in Animations):}}

\begin{itemize}
    \item \textit{\textbf{Taming-VQGAN}} \cite{esser2021taming}: It uses the CNN to extract visual features from the image and generate a codebook of context-rich visual tokens. The transformers are then employed to capture long-range and global interactions among these visual tokens. By combining both components, the model is capable of generating high-resolution images.

    \item \textit{\textbf{OmniTokenizer}}\cite{wang2024omnitokenizer}: OmniTokenizer is a transformer-based unified tokenizer for images and videos, featuring a spatial-temporal decoupled architecture that leverages window and causal attention. Using a progressive training strategy, it efficiently processes both image and video data within a single framework and achieves state-of-the-art reconstruction performance.

    \item \textit{\textbf{LmkCodec-GRVQ}}: A variant of LmkCodec replaces the FSQ layer with GRVQ. Specifically, GRVQ is configured with 4 groups, 2 quantizers, and the codebook size is 1024.

    \item \textit{\textbf{LmkCodec-VQ}}: A variant of LmkCodec replaces the FSQ layer with VQ. Specifically, its codebook size is 1024.
    
\end{itemize}

\subsubsection{\textbf{Second category of baseline models (Synthesizing Speech in Conversational Scenarios):}}

\begin{itemize}
    \item \textit{\textbf{GRU-CSS}} \cite{guo2021conversational}: A GRU-based CSS model that extracts sentence-level representations from the dialogue context to capture inter-sentence dependencies. These contextual features are then fed into a FastSpeech2 \cite{ren2020fastspeechs} backbone to synthesize the target speech.

    \item \textit{\textbf{M$^2$-CTTS}} \cite{xue2023m2ctts}: A multi-scale, multimodal CSS model that designs dedicated context modules for both text and speech to simultaneously capture dialogue history at both coarse and fine granularities, the backbone is fastspeech2.

    \item \textit{\textbf{MSRGCN-CSS}} \cite{li2022inferring}: A CSS model based on a Multi-Scale Relational Graph Convolutional Network (MSRGCN), which captures multimodal dependencies in dialogue at both global (sentence-level) and local (word-level) scales, the backbone is FastSpeech2.

    \item \textit{\textbf{ECSS}} \cite{liu2024emotion}: A CSS model with context modeling based on a heterogeneous graph, which renders the target speech by predicting both emotion categories and emotion intensity. The backbone of the model is FastSpeech2.

    \item \textit{\textbf{GPT-Talker}} \cite{liu2024gpttalker}: A GPT-style CSS model that converts multimodal information from multi-turn dialogue history into token sequences. It uses a GPT-based architecture to predict speech tokens that including both semantic content and speaking style, followed by speech synthesis using an enhanced dialogue-aware version of VITS \cite{kim2021conditional}.
    
    \item \textit{\textbf{Empatheia$^\S$}} \cite{zhang2025empatheia}: A variant of Empatheia that removes visual information from both the model’s inputs and outputs.

     \item \textit{\textbf{EmpathyEar$^\S$}} \cite{fei2024empathyear}:  A variant of EmpathyEar that removes visual information from both the model’s inputs and outputs.
\end{itemize}


\subsubsection{\textbf{Third category of baseline models (Generating Talking-face Animations in Conversational Scenarios):}}
\begin{itemize}
    \item \textit{\textbf{AniPortrait}} \cite{wei2024aniportrait}: It generates high-quality animations by leveraging speech and a reference portrait through a two-stage framework. First, it extracts 3D intermediate representations from the speech and converts them into 2D facial landmark sequences. Then, a diffusion model combined with a motion module transforms the landmark sequences into realistic and temporally consistent talking-face animations.

    \item \textit{\textbf{Hallo}} \cite{xu2024hallo}: It proposes an end-to-end diffusion-based architecture that integrates a hierarchical speech-driven visual synthesis module. And it combines a diffusion-based generative framework, a UNet denoiser, temporal alignment techniques, and a reference network.

    \item \textit{\textbf{Echomimic}} \cite{chen2024echomimic}: It leverages both speech and facial landmarks, combining the strengths of both modalities to achieve more stable and natural portrait video generation. However, since it cannot directly predict the landmarks of the target utterance, only speech is used for driving in experimental comparisons.

    \item \textit{\textbf{SadTalker}} \cite{zhang2023sadtalker}: It generates 3D motion parameters from speech—including head pose and facial expressions—and employs ExpNet and PoseVAE to accurately model expressions and pose separately. These 3D motion parameters are then mapped to 3D-aware facial rendering for realistic animation.

    \item \textit{\textbf{Empatheia}} \cite{zhang2025empatheia}: We implement a variant based on the original model, where the target utterance is fixed. It leverages multimodal context awareness and the Vicuna \cite{chiang2023vicuna} to predict the speech-visual content and style of the target utterance. StyleTTS2 \cite{li2023styletts2} and DreamTalker \cite{ma2023dreamtalk} are then used to synthesize the speech and talking-face animations with style informations, respectively.

    \item \textit{\textbf{EmpathyEar}} \cite{fei2024empathyear}: We implement a variant based on the original model, where the target utterance is fixed. It leverages multimodal context awareness and the Chat-GLM3 \cite{du2021glm} to predict the style of the target utterance. StyleTTS2 \cite{li2023styletts2} and EAT \cite{gan2023eat} are then used to synthesize the speech and talking-face animations  with style informations, respectively.
\end{itemize}

\subsection{Metrics}
\subsubsection{More details on subjective experiments}
In the subjective evaluation, we recruited 30 systematically trained university students, all of whom speak English as a second language and possess strong skills in listening, speaking, reading, and writing. Each participant receives a payment of 20 \$ for completing each group of subjective tests, which is considered fair compensation locally. Before each test session, all participants listen to provided sample speeches in a quiet environment to help them understand the scoring criteria. All evaluations follow the standard MOS (Mean Opinion Score) rules, using a 1–5 scale, where 1 is Bad, 2 is Poor, 3 is Fair, 4 is Good, and 5 is Excellent. The order of samples is randomized in each test to minimize bias in participants’ judgments.

\subsubsection{Instructions for subjective evaluation}
We provide detailed instructions to participants during the subjective evaluation, as shown below.


\tcbset{
  myboxstyle/.style={
    enhanced,
    sharp corners=south,
    colback=white,
    colframe=black,
    boxrule=1pt,
    arc=4pt,
    top=2mm,
    bottom=2mm,
    left=2mm,
    right=2mm,
    fonttitle=\bfseries,
    coltitle=white,
    colbacktitle=black,
    title={Instructions of \textit{MOS$_{SN}$}}
  }
}
\begin{tcolorbox}[myboxstyle]
Please refer to the speech and text of the dialogue history, as well as the text of the current sentence. Focus solely on the naturalness and quality of the speech generated by each model, including correct pronunciation, absence of noise, absence of unreasonable pauses, consistency of timbre, etc. For optimal results, please wear headphones and work in a quiet environment.
\end{tcolorbox}

\tcbset{
  myboxstyle/.style={
    enhanced,
    sharp corners=south,
    colback=white,
    colframe=black,
    boxrule=1pt,
    arc=4pt,
    top=2mm,
    bottom=2mm,
    left=2mm,
    right=2mm,
    fonttitle=\bfseries,
    coltitle=white,
    colbacktitle=black,
    title={Instructions of \textit{MOS$_{SE}$}}
  }
}
\begin{tcolorbox}[myboxstyle]
Please refer to the speech and text of the dialogue history, as well as the text of the current sentence. Focus solely on whether the emotion conveyed by the speech generated by each model aligns with the current dialogue context and whether the expressed emotion can achieve empathy. For optimal results, please wear headphones and work in a quiet environment.
\end{tcolorbox}

\tcbset{
  myboxstyle/.style={
    enhanced,
    sharp corners=south,
    colback=white,
    colframe=black,
    boxrule=1pt,
    arc=4pt,
    top=2mm,
    bottom=2mm,
    left=2mm,
    right=2mm,
    fonttitle=\bfseries,
    coltitle=white,
    colbacktitle=black,
    title={Instructions of \textit{MOS$_{VN}$}}
  }
}
\begin{tcolorbox}[myboxstyle]
Please refer to the speech, text, and talking-face animations of the dialogue history, as well as the text of the current sentence. Focus solely on the quality and naturalness of the generated talking-face animations, including smooth facial movements without jitter, clear image quality, consistency of facial features with the reference image, undistorted facial components, and a stable background throughout. For optimal results, please wear headphones and work in a quiet environment.
\end{tcolorbox}

\tcbset{
  myboxstyle/.style={
    enhanced,
    sharp corners=south,
    colback=white,
    colframe=black,
    boxrule=1pt,
    arc=4pt,
    top=2mm,
    bottom=2mm,
    left=2mm,
    right=2mm,
    fonttitle=\bfseries,
    coltitle=white,
    colbacktitle=black,
    title={Instructions of \textit{MOS$_{VE}$}}
  }
}
\begin{tcolorbox}[myboxstyle]
Please refer to the speech, text, and talking-face animations of the dialogue history, as well as the text of the current sentence. Focus solely on whether the facial expressions are natural and undistorted, whether they align with the emotions conveyed in the generated speech, and whether they are appropriate for the current dialogue context. For optimal results, please wear headphones and work in a quiet environment.
\end{tcolorbox}

\tcbset{
  myboxstyle/.style={
    enhanced,
    sharp corners=south,
    colback=white,
    colframe=black,
    boxrule=1pt,
    arc=4pt,
    top=2mm,
    bottom=2mm,
    left=2mm,
    right=2mm,
    fonttitle=\bfseries,
    coltitle=white,
    colbacktitle=black,
    title={Instructions of \textit{MOS$_{VC}$}}
  }
}
\begin{tcolorbox}[myboxstyle]
Please refer to the speech, text, and talking-face animations of the dialogue history, as well as the text of the current sentence. Focus solely on whether the lip movements match the spoken content. For optimal results, please wear headphones and work in a quiet environment.
\end{tcolorbox}

\section{More Details of Results}
\subsection{More visual comparison results about Talking-face animations}
\begin{figure*}[t]
\centering
\centerline{
\includegraphics[width=1\linewidth]{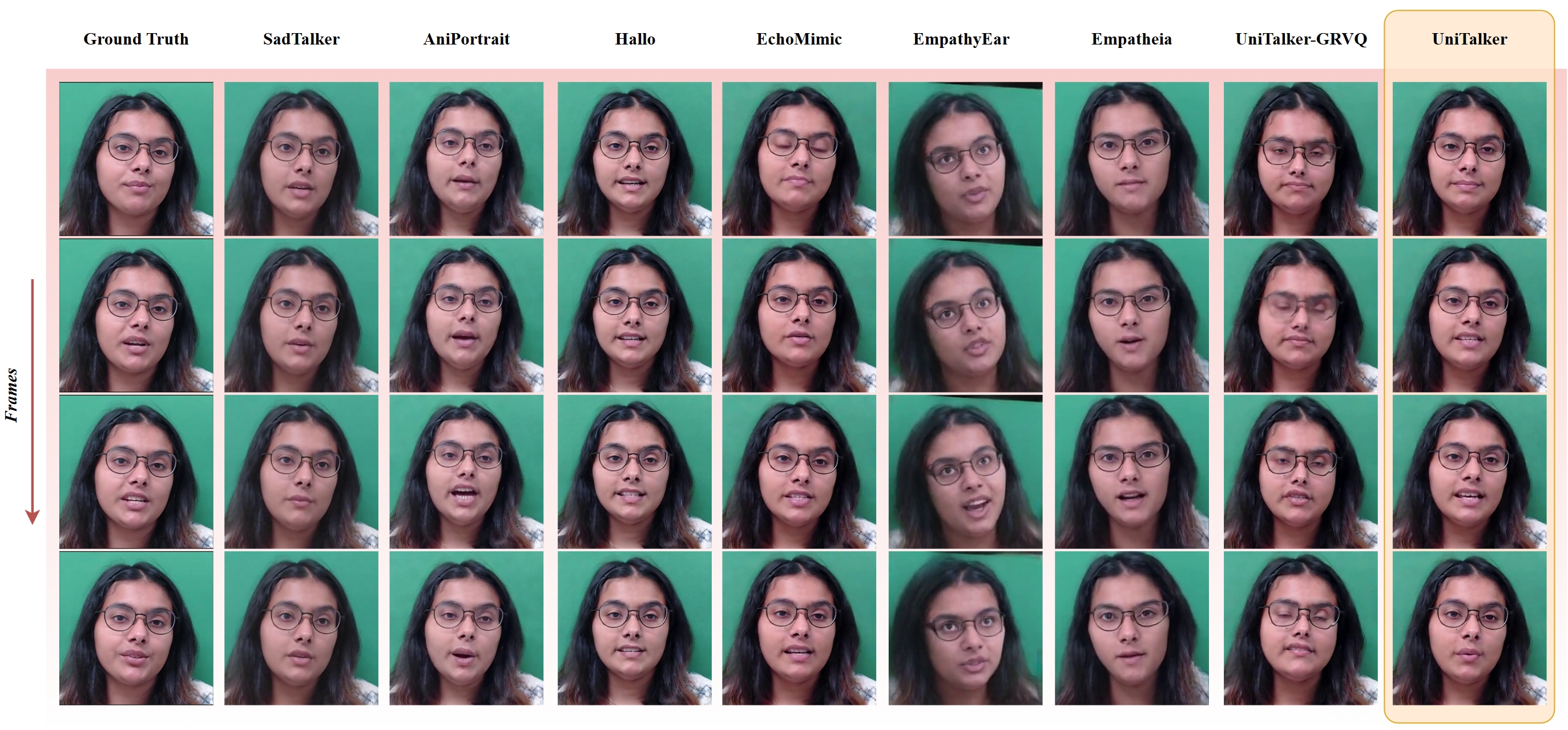}
}
\caption{Visualization of talking-face animations generated by UniTalker and baseline models on the MultiDialog dataset.}
\label{fig:visiual-2}
\end{figure*}

As mentioned in the "Visualization Results" subsection, we additionally include another set of experimental results. As shown in Fig. \ref{fig:visiual-2}, the talking-face animations generated by UniTalker consistently maintain the highest quality.

\subsection{Zero-shot Timbre Rendering Performance Analysis}
Leveraging large-scale speech data training, UniTalker enables zero-shot speaker timbre rendering in the CSVS task. To evaluate this capability, we conduct a speaker similarity experiment using 100 dialogue samples from four unseen speakers in the IEMOCAP \cite{busso2008iemocap} dataset. Each CSVS model is provided with the dialogue history, along with the current speaker, text, facial landmarks, and emotion label, and tasked with synthesizing the corresponding target speech. As reported in Table~\ref{tab:appendix-2}, UniTalker achieves the highest speaker similarity among all compared models. Although its performance is slightly lower than the results obtained on the MultiDialog dataset (Table~\ref{tab:exp-3}), this is likely attributable to the presence of noise and low-quality facial imagery in IEMOCAP. Nonetheless, these results demonstrate that UniTalker possesses fundamental zero-shot speaker timbre cloning capability.

\subsection{Case Study}
\subsubsection{\textbf{Examples of UniTalker generating talking-face animations with different emotions}}
\begin{figure*}[t]
\centering
\centerline{
\includegraphics[width=0.98\linewidth]{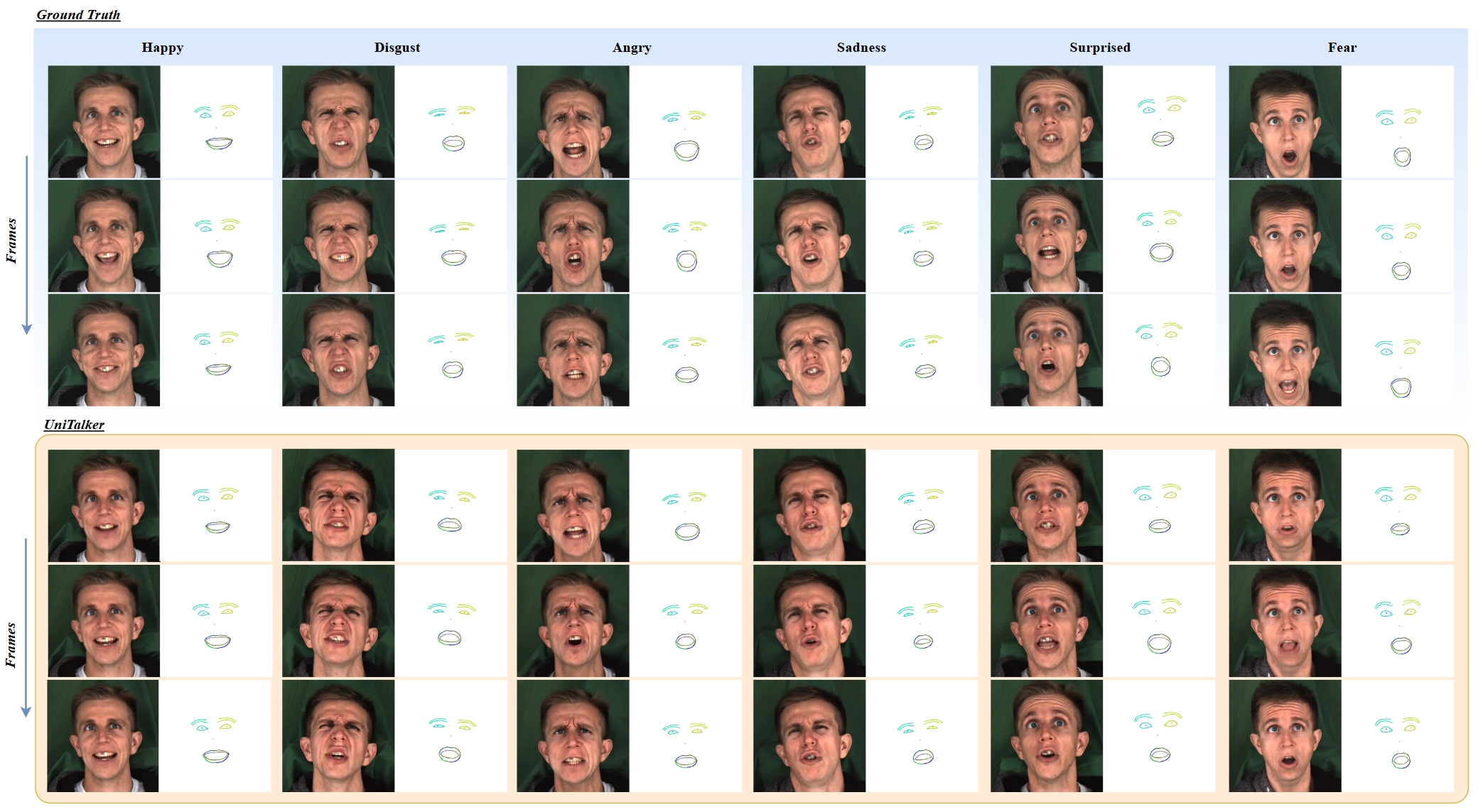}
}
\caption{Visualization of UniTalker’s talking-face animations generation across different emotions (speaker A).}
\label{fig:visiual-3}
\end{figure*}

\begin{figure*}[t]
\centering
\centerline{
\includegraphics[width=0.98\linewidth]{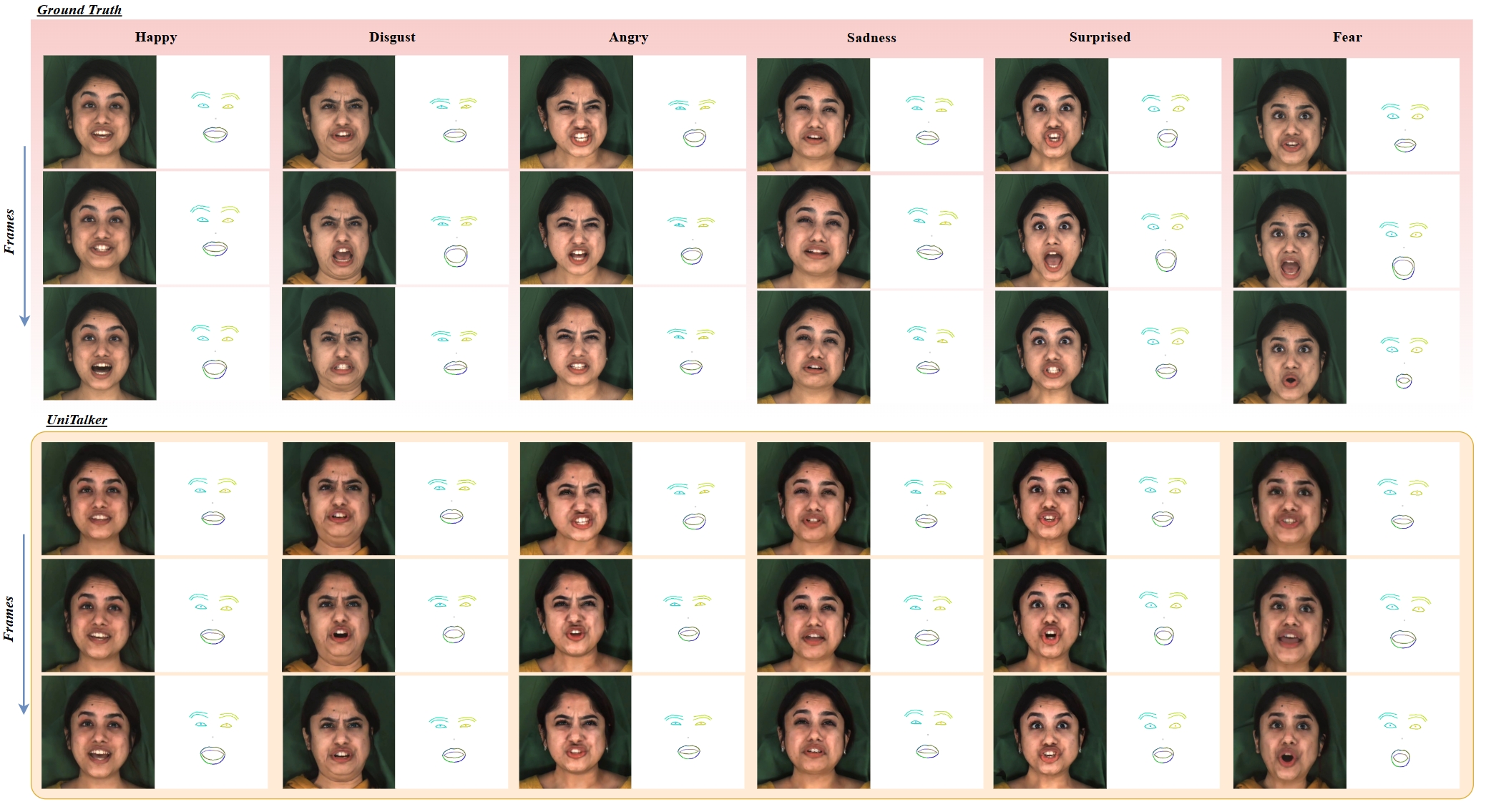}
}
\caption{Visualization of UniTalker’s talking-face animations generation across different emotions (speaker B).}
\label{fig:visiual-4}
\end{figure*}

To intuitively demonstrate UniTalker's ability to render visual responses for different emotional categories, we select two speakers from the MEAD \cite{wang2020MEAD} dataset. For each experiment, we specify the speaker, text, emotion label, and speech as input to the model (the speaker is extracted from the given speech and the reference portrait is the first frame image of Ground Truth). UniTalker then predicts the facial landmarks of the target utterance and renders the corresponding talking-face animations. As shown in Fig.~\ref{fig:visiual-3} and Fig.~\ref{fig:visiual-4}, compared with the Ground Truth, UniTalker generates distinct facial expressions aligned with different emotions while effectively preserving the speaker’s identity and background information.

\subsubsection{\textbf{Examples of UniTalker's zero-shot ability to generate talking-face animations for unseen speakers}}
\begin{figure*}[t]
\centering
\centerline{
\includegraphics[width=0.98\linewidth]{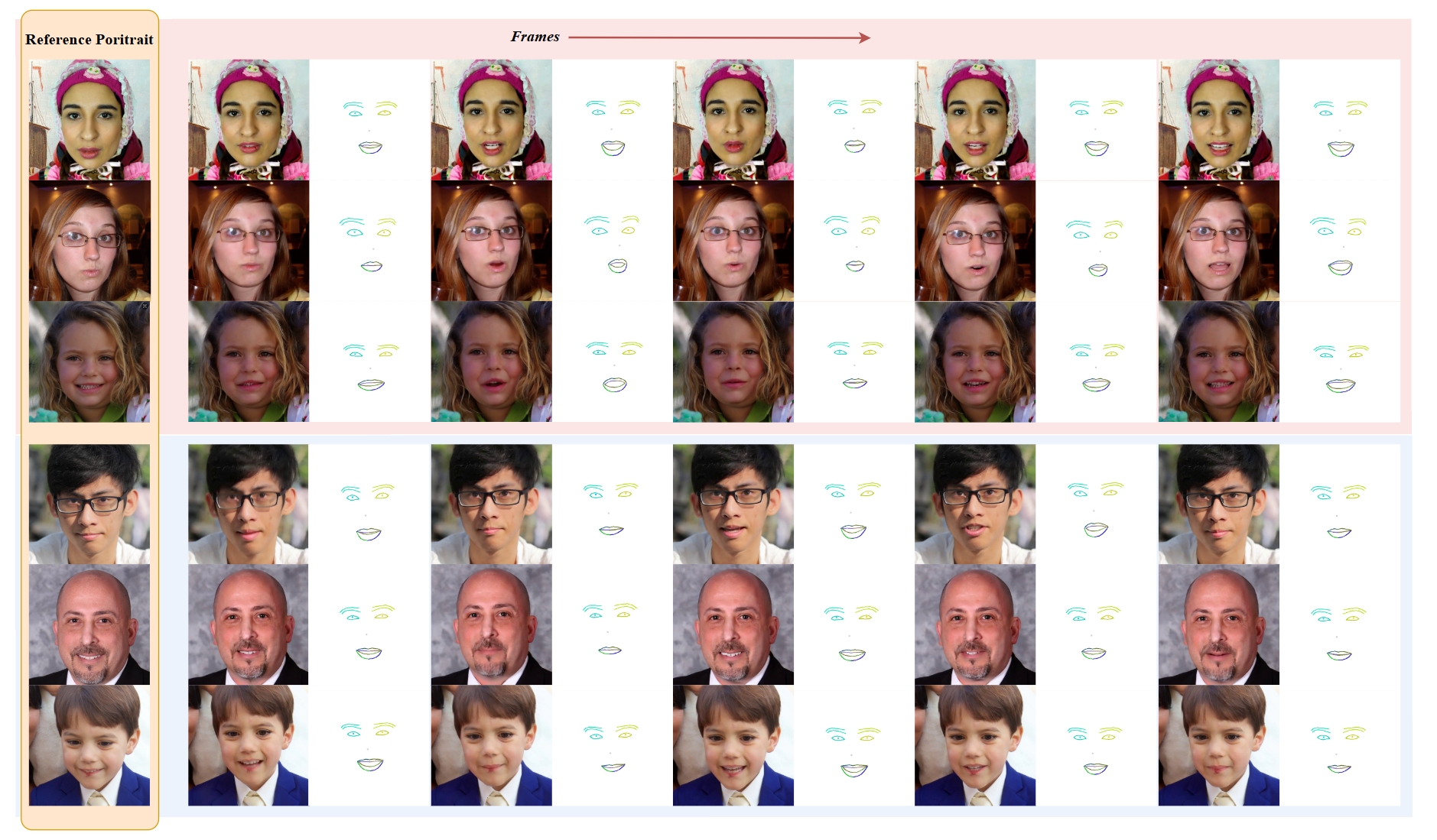}
}
\caption{Visualization of UniTalker's zero-shot talking-face animations generation for unseen speakers.}
\label{fig:visiual-5}
\end{figure*}

To further explore UniTalker's zero-shot ability in talking-face animation generation, we conduct two group experiments using speakers from the FFHQ \cite{karras2019ffhq} dataset who were not seen during training. Specifically, we select three individuals of both genders as reference portraits. In each experiment, the same speaker, text, emotion label, and speech are provided as input (the speaker identification is extracted from the given speech), and UniTalker predicts the facial landmarks of the target utterance, followed by talking-face animation rendering. As shown in Fig.~\ref{fig:visiual-5}, the generated animations appear natural, without any noticeable facial distortion. Additionally, since each experimental group is provided with the same speech, the speakers in each group maintain consistent lip movements.

\section{Limitations and Future Work}
We evaluate the performance of UniTalker on a local machine equipped with an NVIDIA GeForce RTX 4080 GPU and 32 GB of RAM. On average, the generation of speech responses completes within approximately 2 seconds, and the rendering of talking-face animations takes about 5 seconds for every 25 frames. While the current animation rendering process requires relatively more time, this latency is acceptable for offline or semi-real-time applications. In future work, we plan to further optimize the efficiency of each module and explore streaming-based approaches to reduce response latency. Additionally, UniTalker currently generates animations at a resolution of 512×512. We consider exploring higher-resolution generation in future work to enhance visual quality.

\begin{table}[t]
\caption{\label{tab:appendix-2} Results of zero-shot timbre rendering experiment.}
\centering
\resizebox{0.8\linewidth}{!}{
\begin{tabular}{@{}c|ccc@{}}
\toprule
\textbf{Models} & \textbf{EmpathyEar} & \textbf{Empatheia} & \textbf{UniTalker} \\ \midrule
SIM$_{SPK}$ ($\uparrow$) & 0.612 & 0.634 & 0.702 \\ \bottomrule
\end{tabular}
}
\end{table}

\end{document}